\documentclass[12pt]{article}
\usepackage{epsfig}
\usepackage{amssymb}
\usepackage{amsmath}
\usepackage{amsfonts}
\usepackage{graphicx}
\usepackage{mathrsfs}
\usepackage[dvips]{color}
\usepackage{multirow}

\newcommand{\R}{\mathbb{R}}
\newcommand{\C}{\mathbb{C}}

\newcommand{\fg}{\mathfrak{g}}
\newcommand{\fh}{\mathfrak{h}}

\newcommand{\fu}{\mathfrak{u}}

\newcommand{\fs}{\mathfrak{s}}
\newcommand{\bfs}{{\boldsymbol{\mathfrak{s}}}}

\newcommand{\fH}{\mathfrak{H}}

\newcommand{\fO}{\mathfrak{O}}

\newcommand{\bg}{\mathbf{g}}

\newcommand{\bh}{\mathbf{h}}

\newcommand{\bs}{{\mathbf{s}}}

\newcommand{\bA}{\mathbf{A}}

\newcommand{\bD}{\mathbf{D}}

\newcommand{\bF}{\mathbf{F}}
\newcommand{\bG}{\mathbf{G}}
\newcommand{\bcG}{\boldsymbol{\cG}}
\newcommand{\bH}{\mathbf{H}}
\newcommand{\bI}{\mathbf{I}}

\newcommand{\bM}{\mathbf{M}}

\newcommand{\bU}{\mathbf{U}}

\newcommand{\balpha}{{\boldsymbol{\alpha}}}
\newcommand{\bsigma}{{\boldsymbol{\sigma}}}

\newcommand{\bomega}{{\boldsymbol{\omega}}}
\newcommand{\bfeta}{{\boldsymbol{\eta}}}
\newcommand{\brho}{{\boldsymbol{\rho}}}
\newcommand{\btheta}{{\boldsymbol{\theta}}}
\newcommand{\bGamma}{{\boldsymbol{\Gamma}}}
\newcommand{\cA}{{\mathcal{A}}}

\newcommand{\cC}{\mathcal{C}}

\newcommand{\cH}{\mathcal{H}}
\newcommand{\cE}{\mathcal{E}}

\newcommand{\cG}{\mathcal{G}}

\newcommand{\cO}{\mathcal{O}}
\newcommand{\cP}{\mathcal{P}}

\newcommand{\cS}{\mathcal{S}}
\newcommand{\cT}{\mathcal{T}}
\newcommand{\cU}{\mathcal{U}}
\newcommand{\bcU}{{\boldsymbol{\cU}}}

\newcommand{\cX}{\mathcal{X}}

\newcommand{\be}{\begin{equation}}
\newcommand{\ee}{\end{equation}}
\newcommand{\bea}{\begin{eqnarray}}
\newcommand{\eea}{\end{eqnarray}}
\newcommand{\nn}{\nonumber}
\newcommand{\kt}{\rangle}
\newcommand{\br}{\langle}

\newcommand{\ed}{\end{document}}

\newcommand{\bi}{\begin{itemize}}
\newcommand{\ei}{\end{itemize}}

\newcommand{\bce}{\begin{center}}
\newcommand{\ece}{\end{center}}

\newcommand{\sH}{\mathscr{H}}

\newcommand{\sP}{\mathscr{P}}

\newcommand{\sT}{\mathscr{T}}

\newcommand{\bzero}{\mathbf{0}}

\begin{document}

\title{Energy Observable for a Quantum System with a Dynamical Hilbert Space and a Global Geometric  Extension of
Quantum Theory\footnote{Dedicated to the memory Bryce and Cecile DeWitt.}}

\author{Ali~Mostafazadeh\thanks{E-mail address: amostafazadeh@ku.edu.tr}~
\\ Departments of Mathematics 
and Physics, 
Ko\c{c} University,\\ 34450 Sar{\i}yer,
Istanbul, Turkey}

\date{ }
\maketitle

\begin{abstract}
A non-Hermitian operator may serve as the Hamiltonian for a unitary
quantum system, if we can modify the Hilbert space of state vectors
of the system so that it turns into a Hermitian operator. If this
operator is time-dependent, the modified Hilbert space is generally
time-dependent. This in turn leads to a generic conflict between the
condition that the Hamiltonian is an observable of the system and
that it generates a unitary time-evolution via the standard
Schr\"odinger equation. We propose a geometric framework for
addressing this problem. In particular we show that the Hamiltonian
operator consists of a geometric part, which is determined by a
metric-compatible connection on an underlying Hermitian vector
bundle, and a non-geometric part which we identify with the energy
observable. The same quantum system can be locally
described using a time-dependent Hamiltonian that acts in a
time-independent state space and is the sum of a geometric part and
the energy operator. The full global description of the system is
achieved within the framework of a moderate geometric
extension of quantum mechanics where the role of the Hilbert
space of state vectors is played by a Hermitian vector bundle $\cE$
endowed with a metric compatible connection, and observables are
given by global sections of a real vector bundle that is determined
by $\cE$. We examine the utility of our proposal to describe a class
of two-level systems where $\cE$ is a Hermitian vector bundle
over a two-dimensional sphere.\\[12pt]


\end{abstract}

\section{Introduction}

The unexpected observation that certain complex  $\cP\cT$-symmetry
potentials \cite{Bender-1998}, such as $v(x)=ix^3$, can have a real
spectrum has motivated many researchers to seek means for employing
them in quantum mechanics. The reality of the spectrum of these
potentials was initially associated with their exact
$\cP\cT$-symmetry\footnote{This means the existence of a complete
basis of square-integrable functions that are common eigenfunctions
of both the Schr\"odinger operator $-\partial_x^2+v(x)$ and
$\cP\cT$, where $\cP$ and $\cT$ stand for the parity and
time-reversal operators; $\cP\psi(x):=\psi(-x)$ and
$\cT\psi(x):=\psi(x)^*$.}. This was later shown to be equivalent to
the Hermitizability of their Schr\"odinger operator,
$H=-\partial_x^2+v(x)$, i.e., the fact that $H$ turns into a
Hermitian operator\footnote{Following von Neumann
\cite{von-Neumann}, we use the terms ``Hermitian'' and
``self-adjoint'' synonymously. For a precise definition see
\cite{review}.} upon an appropriate modification of the Hilbert
space it acts in \cite{p2,jpa-2003}. Indeed if we identify the
mean value of the outcomes of measurements of physical
observables with the standard expression for the expectation
values of a linear operator $O$, i.e., $\frac{\br\psi,O\psi\kt}{\br\psi,\psi\kt}$,
then according to a mathematical result, that is unfortunately not so well-known among
physicists, the requirement that the expectation values must be real
numbers implies that the operator must be Hermitian~\cite{review,cjp-2004}. In short the reality of expectation values implies the Hermiticity of the operator, while the reality of
the spectrum does not.

Although the study of $\cP\cT$-symmetric  Hamiltonians did not
actually lead to a genuine extension of quantum mechanics \cite{cjp-2004},
as was initially claimed \cite{bender-prl-2002},
it revealed the existence of alternative representations of quantum mechanics. These
subsequently found applications in such areas as quantum
cosmology, relativistic quantum mechanics, and classical
electrodynamics \cite{review}. Another important development that was
triggered by the study of $\cP\cT$-symmetry is the consequences of
its realization and applications in classical optics
\cite{muga,PT-optics1,PT-optics2,PT-optics3,PT-optics4,PT-optics5}.
For a recent review, see \cite{konotop-RMP}.

The above-mentioned Hermitization procedure applies to non-Hermitian
operators $H$ that satisfy the $\eta$-pseudo-Hermiticity condition
\cite{p1}:
    \be
    H^\dagger=\eta H\eta^{-1}.
    \label{ph}
    \ee
Here $H^\dagger$ is the adjoint of $H$, $\eta$ is a bounded  and
inversely bounded positive-definite operator called the metric
operator \cite{review}, and $H$ and $\eta$ are assumed to act in an
auxiliary Hilbert space $\sH$. For the cases that $H$ has a real
discrete spectrum and a complete set of eigenvectors forming a
(Riesz) basis of $\sH$, metric operators satisfying (\ref{ph}) have
the form $\eta=\sum_n|\phi_n\kt\br\phi_n|$, where $\phi_n$ are
eigenvectors of $H^\dagger$  that together with a set of
eigenvectors $\psi_n$ of $H$ form a complete biorthonormal system
\cite{p1}. If (\ref{ph}) holds, one can modify $\sH$ by endowing the
vector space of its elements with the inner product defined by
$\eta$, namely
    \be
    \br\cdot,\cdot\kt_{\eta}:=\br\cdot|\eta\cdot\kt,
    \label{inn-prod}
    \ee
where $\br\cdot|\cdot\kt$ stands for the inner product of $\sH$, \cite{p2}.

The choice of $\eta$ is not unique. Each choice determines a
corresponding Hilbert space $\sH_{\eta}$ in which $H$ acts as a
Hermitian operator; $\br\cdot,H \cdot\kt_{\eta}=\br
H\cdot,\cdot\kt_{\eta}$. The pair $(\sH_{\eta},H)$ determines a
quantum system whose obervables are represented by the Hermitian
operators acting in $\sH_{\eta}$ and whose dynamics is determined by
the Schr\"odinger equation:
    \be
    i\frac{d}{dt}\psi(t)=H\psi(t).
    \label{sch-eq}
    \ee
The same quantum system can be represented by the Hilbert space and
Hamiltonian pair:  $(\sH,h)$, where $h:\sH\to\sH$ is the Hermitian
operator
    \be
    h:=\rho\,H\rho^{-1},
    \label{h=}
    \ee
and $\rho$ is the positive square root  of $\eta$, \cite{jpa-2003}.
In practice, the only reason one might want to describe the quantum
system using its pseudo-Hermitian representation, $(\sH_{\eta},H)$,
is the complicated and often nonlocal nature of $h$. See for example
\cite{jpa-2015}.

In implementing the approach we just outlined, one faces two major
difficulties. The first stems from the technical difficulties of
dealing with operators acting in an infinite dimensional Hilbert
space. In particular, the familiar choices of $\sH$ might obstruct
the existence of a bounded and inversely bounded metric operator
satisfying (\ref{ph}), \cite{david-petr}. The second is related to a
no-go theorem regarding the conflict between the pseudo-Hermiticity
relation~(\ref{ph}) for a time-dependent Hamiltonian operator,
$H=H(t)$, which identifies it with an observable, and the
requirement of the unitarity of the dynamics determined by the
Schr\"odinger equation (\ref{sch-eq}) in the physical Hilbert space
$\sH_{\eta}$, \cite{plb-2007}. Time-dependent pseudo-Hermitian
Hamiltonians that admit time-independent metric operators were
originally  considered in Ref.~\cite{Fring-2006}. Pseudo-Hermitian
Hamiltonians that require dealing with time-dependent metric
operators have been studied in
Refs.~\cite{cqg-2003,ap-2004,plb-2007,znojil-2008,cong-2013,znojil-2015,Fring-2016,Fring-2017,Fring-2017b,znojil-2017,Fring-2018a,Fring-2018b,znojil-2018}.
They appear naturally in connection with the Hilbert-space problem
in quantum cosmology \cite{cqg-2003,ap-2004}.

Ref.~\cite{ptrsa-2013} proposes a  solution for the problem of the
nonexistence of bounded and inversely bounded metric operators for
the case that $H$ has a real and discrete spectrum. This involves a
minimal modification of $\sH$ as a set such that the modified Hilbert
space contains the span of eigenvectors of $H$ as a dense subspace.

The second problem seems to have two  alternative solutions
\cite{plb-2007}; one should either modify the Schr\"odinger equation
(\ref{sch-eq}) as suggested by \cite{znojil-2008}, or accept that
$H(t)$ is not an observable, i.e., it is a generator of dynamics
that differs from the energy operator \cite{Fring-2016}.
This in turn leads to another difficulty, namely lack of a physical
or mathematical evidence for superiority of one of these choices
over the other. In the present paper we offer a geometric framework
that elucidates the conceptual basis of this problem and offers a
satisfactory solution for it. This in turn reveals the geometric
meaning of the energy operator and provides a natural route towards
a geometric extension of quantum mechanics.

We begin our discussion by recalling the basic argument leading to
the no-go theorem of Ref.~\cite{plb-2007}.

Consider a time-dependent Hamiltonian  operator $H=H(t)$ acting in a
physical Hilbert space $\sH_{\eta(t)}$ that is determined by a
time-dependent metric operator $\eta(t)$. The unitarity of dynamics
means that the inner product of any pair of evolving state vectors,
$\phi(t)$ and $\psi(t)$, is time-independent;
    \be
    \frac{d}{dt}\:\br\phi(t),\psi(t)\kt_{\eta(t)}=0.
    \label{e1}
    \ee
Let $t_0\in\R$ mark the initial time, and $U(t,t_0)$ be the associated time-evolution
operator, so that
    \be
    \psi(t)=U(t,t_0)\psi(t_0),~~~~t_0,t\in\R.
    \label{U=}
    \ee
We can use this relation to write the Schr\"odinger equation~(\ref{sch-eq}) in the form
    \be
    i\frac{d}{dt} U(t,t_0)=H(t)U(t,t_0).
    \label{sch-eq-U}
    \ee
In light of (\ref{inn-prod}) and (\ref{sch-eq-U}),  (\ref{e1}) takes the form
    \be
    H^\dagger-\eta H\eta^{-1}=i\dot\eta\,\eta^{-1},
    \label{eq2}
    \ee
where we suppress the time-dependence of $H$ and $\eta$  for brevity
and use an overdot to denote differentation with respect to $t$. It
is clear from (\ref{eq2}) that unless $\eta$ is time-independent,
$H$ does not satisfy the pseudo-Hermiticity relation (\ref{ph}).
This in turns means that it does not act as a Hermitian operator in
the physical Hilbert space $\sH_{\eta(t)}$. Hence, it does not define an
observable.

As one can easily see from (\ref{eq2}), it is the time-dependence of
the metric operator that is responsible for the apparently
undesirable conclusion that the Hamiltonian operator fails to be an
observable. This calls for a deeper investigation of the
consequences of the time-dependence of the metric operator and the
corresponding physical Hilbert space. Indeed, the same conceptual
problem arises in more general situations where the state space of a
quantum system undergoes temporal changes. A typical example is a
particle constrained to move on a surface (or more generally a
Riemannian manifold) with a dynamical shape (respectively geometry)
\cite{DeWitt-1957}. This is indeed a simple example of a physical
system with a dynamical background. The study of such systems has
served as a starting point for attempts towards quantization of
classical fields defined in curved spacetimes and the more basic
problem of the quantization of gravity. The development of quantum
field theories in nonstationary \cite{BD,fullings,DeWitt-QFT} (in
particular cosmological \cite{rubakov-1984}) backgrounds involves
dealing with time-dependent state spaces. Another problem in which
time-dependent state spaces and metric operators play a basic role
is in the approach proposed in Ref.~\cite{ap-2004} for dealing with
the Hilbert-space problem in minisuperspace quantum cosmology.

Because the finite/infinite-dimensionality of the Hilbert space is
of no relevance to the difficulty with unobservability of the Hamiltonian
operator for a unitary system with a time-dependent state space, in what follows
we confine our attention to systems with a finite-dimensional Hilbert space.

\section{Quantum systems with a time-dependent state space}
\label{S2}

Consider a quantum system whose kinematical features are described by a time-dependent Hilbert space $\cH_t$ with a finite and constant dimension $N$. Suppose that $\cH_t$ is obtained by endowing a complex vector space $V_t$ with an inner product $\br\cdot,\cdot\kt_{t}$. The dynamics of state vectors of the system should be derived from a rule for computing the rate of change of the state vectors in time. In a period of time, $[t,t+\epsilon]$, a state vector $\psi(t)$ changes to $\psi(t+\epsilon)$. Therefore it is tempting to identify the rate of change of $\psi(t)$ at $t$ with its time-derivative:
    \[\lim_{\epsilon\to 0} \frac{\psi(t+\epsilon)-\psi(t)}{\epsilon}.\]
This is however problematic, because $\psi(t)$ and $\psi(t+\epsilon)$ belong to different vector spaces, namely $V_t$ and $V_{t+\epsilon}$, and it is not meaningful to add or subtract vectors belonging to different vector spaces.

What one can do is to find a prescription to associate a unique element of $\cH_t$ to $\psi(t+\epsilon)$, say $\psi_\epsilon(t)$, and quantify the rate of change of $\psi(t)$ with
    \be
    \lim_{\epsilon\to 0}\frac{\psi_\epsilon(t)-\psi(t)}{\epsilon}.
    \label{cov-der-t}
    \ee
The prescription that determines $\psi_\epsilon(t)$ is called a connection or parallel transportation, and (\ref{cov-der-t}) is called the covariant time-derivative associated with this connection. The proper mathematical tools for describing this phenomenon is provided by the theory of vector bundles \cite{kobayashi} which is widely used in the study of gauge theories \cite{cecile} and geometric phases \cite{book2}.

If the time-dependence of $\cH_t$ is induced by changing a finite number of relevant physical parameters: $R^1,R^2,\cdots,R^d$, which we collectively denote by $R$, we can identify the vector space of state vectors, the inner product that promotes it to a Hilbert space, and the resulting Hilbert space for each value of $R$ respectively by $V_R$, $\br\cdot,\cdot\kt_R$, and $\cH_R$. This allows us to consider different dynamical changes of the state space of the system by considering different parameterized curves $\cC:[t_1,t_2]\to M$, where $M$ stands for the parameter space of the system whose points are labelled by $R$. In general $M$ is a smooth manifold \cite{cecile} and $R$ provides the coordinates of its points in a local coordinate patch $\cO_\alpha$. We identify the points of $M$ with their local coordinates $R$ unless otherwise is clear from the context. In particular, we use $R(t)$ to denote the local coordinates of the point $\cC(t)$ on a segment of $\cC$ that lies in $\cO_\alpha$. Clearly, $V_t=V_{R(t)}$ and $\br\cdot,\cdot\kt_{t}=\br\cdot,\cdot\kt_{R(t)}$.

Next, we consider a smooth complex vector bundle $\cE$ over the base manifold $M$ and identify $V_R$ with the fiber of $\cE$ over the point $R\in M$.\footnote{For a friendly introduction to vector bundles, see \cite{book2}.} The typical fiber $V$ of $\cE$ is therefore a complex $N$-dimensional vector space which we can identify with $\C^N$. Let $\pi:\cE\to M$ be the projection map of $\cE$, so that $V_R$ is the inverse image of $\{R\}$ under $\pi$;
    \[V_R=\pi^{-1}(\{R\})=\{~p\in\cE~|~\pi(p)=R~\}.\]
By the very definition of a smooth vector bundle, we can cover $M$ using a collection of its local coordinate patches, $\cO_\alpha$, and there are smooth diffeomorphisms\footnote{A diffeomorphism is a smooth everywhere-defined one-to-one and onto function (i.e., a smooth bijection) with a smooth inverse.} $f_\alpha:\pi^{-1}(\cO_\alpha)\to\cO_\alpha\times V$  that map the fibers of the bundle to its typical fiber isomorphically, i.e., for all $R\in\cO_\alpha$ there is a vector-space isomorphism $\varphi_{\alpha,R}:V_R\to V$ depending smoothly on $R$ such that for each $\psi_R\in V_R$, $f_\alpha(\psi_R)=\big(R,\varphi_{\alpha,R}(\psi_R)\big)$. The pairs $(\cO_\alpha,f_\alpha)$ and the functions
    \be
    g_{\alpha\beta,R}:=\varphi_{\alpha,R}\circ \varphi_{\beta,R}^{-1}:V\to V,
    \label{transition-fns}
    \ee
which are defined for all $R\in\cO_\alpha\cap\cO_\beta$, are respectively called the local trivializations and transition functions of $\cE$. The vector bundle can be viewed as the collection of all $\cO_\alpha\times V$ that are glued along the intersections of $\cO_\alpha$'s according to the following rule: For all $\alpha$ and $\beta$, if $R\in\cO_\alpha\cap\cO_\beta$, then each element $(R,v)$ of $\cO_\alpha\times V$ is identified with (glued to) the element $(R,g_{\beta\alpha,R}(v))$ of $\cO_\beta\times V$.

The fact that the fibers $V_R$ of $\cE$ are endowed with an inner product $\br\cdot,\cdot\kt_R$ shows that the vector bundle $\cE$ is equipped with a Hermitian metric structure \cite{kobayashi}. This in trun implies the existence of a metric-compatible connection $\cA$ on $\cE$. In order to arrive at a local description of such a connection, we choose an orthonormal basis $\{\psi_n[R]\}$ of $\cH_R$ with a smooth dependence on $R$. This defines a set of local sections $\psi_n:\cO_\alpha\to\cE$ that provide a local description of the global sections $\psi:M\to\cE$ of $\cE$ according to
    \be
    \psi[R]=\sum_{n=1}^N\Psi_n[R]\psi_n[R].
    \label{eq12}
    \ee
Here $R\in\cO_\alpha$, and $\Psi_n:\cO_\alpha\to\C$ are smooth functions which we call the components of $\psi$ in the basis $\{\psi_n\}$.

A connection $\cA$ on $\cE$ can be locally described by the following prescription for infinitesimal parallel transportation of vectors along $\cC$: A generic vector belonging to $V_{R(t)}$, which we can express as $\psi[R(t)]=\sum_{n=1}^N\Psi_n[R(t)]\psi_n[R(t)]$, is transported to:
    \be
    \psi(t+dt):=\sum_{m=1}^N\Psi_m(t+dt)\,\psi_m[R(t+dt)]\in V_{R(t+dt)},
    \ee
where
    \be
    \Psi_m(t+dt):=\Psi_m[R(t)]-i
    \sum_{n=1}^N
    \sum_{a=1}^d A_{amn}[R(t)]\,dR^a(t)\Psi_n[R(t)],
    \ee
and $A_{amn}:\cO_\alpha\to\C$ are smooth functions.
The one-forms $A_{mn}:=\sum_{a=1}^d
A_{amn}[R]dR^a$ are the entries of a matrix-valued one-form $\bA$ known as a local connection one-form on $\cE$. The parallel transportation of a vector $\psi(t_0)\in V_{R(t_0)}$ along an extended segment of $\cC$ that lies in $\cO_\alpha$ is determined by identifying the components $\Psi_m(t)$ of the parallel-transported vector $\psi(t)$ by the solution of the initial-value problem:
    \bea
    &&i\frac{d}{dt}\Psi_m(t)=\sum_{n=1}^N \sum_{a=1}^d
    \dot R^a(t) A_{amn}[R(t)]\Psi_n[R(t)],
    \label{eq4}\\
    &&\Psi_m(t_0)=\Psi_m[R(t_0)].
    \eea

In terms of the column vector $\boldsymbol{\Psi}(t)$ which has $\Psi_m(t)$ as its entries, (\ref{eq4}) takes the form
    \be
    \bD_t\boldsymbol{\Psi}(t)=\bzero,
    \label{con-der}
    \ee
where
    \be
    \bD_t:=\frac{d}{dt}+i \sum_{a=1}^d\dot R^a(t)\bA_a[R(t)]
    \label{cov-der=}
    \ee
is the local representation of the covariant time-derivative defined by the connection $\cA$, $\bA_a[R]$ are the $N\times N$ matrices with entries $A_{amn}[R]$, and $\bzero$ is the $N\times 1$ null matrix.

We can express the solution of (\ref{con-der}) in the form
    \be
    \boldsymbol{\Psi}(t)=\bG[R(t)]\boldsymbol{\Psi}(t_0)
    \label{eq5}
    \ee
where
    \be
    \bG[R(t)]:=\sT\exp\left\{-i\int_{t_0}^t \sum_{a=1}^d\bA_a[R(t)] \dot R^a(t) dt\right\}=
    \sP\exp\left\{-i\int_{R(t_0)}^{R(t)}\bA[R]\right\},
    \label{eq6}
    \ee
$\sT$ is the time-ordering operation, and $\sP$ is the path-ordering operation associated with the segment of the curve $\cC$ that connects $R(t_0)$ to $R(t)$. Notice also that setting $\bA_a(t):=\bA_a[R(t)]$ and $\bG(t):=\bG[R(t)]$, we have
    \be
    i\frac{d}{dt} \bG(t)= \sum_{a=1}^d \dot R^a(t)\bA_a(t)\bG(t).
    \label{eq7}
    \ee

We may view a connection $\cA$ on $\cE$ as a rule that assigns to each $\cO_\alpha$ a local connection one-form $\bA$ in such a way that in the intersection of any two local patches, say $\cO_\alpha$ and $\cO_\beta$, the rule for parallel transportation in both of these patches produces the same expression for the parallel-transported vector $\psi(t)$.  This imposes the following transformation rule for the local connection one-forms \cite{cecile}:
    \be
    \bA[R]\to \tilde{\bA}[R]=\bg_{\alpha\beta,R}^{-1}\,\bA[R]\,\bg_{\alpha\beta,R}
    -i\bg_{\alpha\beta,R}^{-1}\,d\bg_{\alpha\beta,R},
    \label{connection-trans}
    \ee
where $\bA$ and $\tilde\bA$ are respectively local connection one-forms associated with the local trivializations $(\cO_\alpha,f_\alpha)$ and $(\cO_\beta,f_\beta)$, $R\in\cO_\alpha\cap\cO_\beta$, $\bg_{\alpha\beta,R}$ is the matrix representation of the transition function $g_{\alpha\beta,R}:\C^N\to\C^N$ in the standard basis of $\C^N$, namely $\{e_1,e_2,\cdots,e_N\}$ with
    \[e_n:=(\,\underbrace{0\,,\,0\,,\,\cdots,\,0}_{ n-1 }\,,\,1\,,\,
    \underbrace{0\,,\,0\,,\,\cdots\,,\,0}_{N-n}\,),\]
$d\bg_{\alpha\beta,R}:=\sum_{a=1}^d\partial_a\bg_{\alpha\beta,R}\,dR^a$, and $\partial_a$ stands for partial differentiation with respect to $R^a$. Note that $\bg_{\alpha\beta,R}$ is the $N\times N$ matrix with entries $\br e_m| g_{\alpha\beta,R}\, e_n\kt$, where $\br\cdot|\cdot\kt$ denotes the Euclidean inner product on $\C^N$; for all $w=(w_1,w_2,\cdots,w_N)\in\C^N$ and $z=(z_1,z_2,\cdots,z_N)\in\C^N$,
    \be
    \br w|z\kt:=\sum_{n=1}^Nw_n^*z_n.
    \label{euclidean}
    \ee

Next, we determine the local connection one-forms $\bA$ associated with a connection $\cA$ that is compatible with the Hermitian metric structure on $\cE$. By definition, these induce parallel transportations that do not change the inner product of the transported vectors along any curve \cite{kobayashi}. To explore the consequences of this condition, consider a pair of arbitrary global sections $\phi,\psi:M\to \cE$. Let $\Phi_n$ and $\Psi_n$ be components of $\phi$ and $\psi$ in the basis $\{\psi_n\}$. Then for all $R\in\cO_\alpha$,
    \be
    \br\phi[R],\psi[R]\kt_R=\sum_{m,n=1}^N \Phi_m[R]^*\eta_{mn}[R]\Psi_n[R]=
    \boldsymbol{\Phi}[R]^\dagger\boldsymbol{\eta}[R]\boldsymbol{\Psi}[R],
    \label{inn-prod-local}
    \ee
where
    \be
    \eta_{mn}[R]:=\br\psi_m[R],\psi_n[R]\kt_R,
    \label{eta-mn}
    \ee
$\dagger$ stands for the conjugate-transpose (Hermitian conjugate) of a matrix, and $\boldsymbol{\eta}[R]$ is the $N\times N$ matrix having $\eta_{mn}[R]$ as its entries. The latter is the local matrix representation of the Hermitian (metric) structure on $\cE$ in the local basis $\{\psi_n\}$.

We can use (\ref{eq5}) and (\ref{inn-prod-local}) to evaluate the inner product of a parallel transported pair of vectors $\phi(t_0),\psi(t_0)\in \cH_{R(t_0)}$ along the segment of the curve $\cC$ that lies in $\cO_\alpha$. This results in
    \be
    \br\phi(t),\psi(t)\kt_{R(t)}=
    \boldsymbol{\Phi}(t)^\dagger\boldsymbol{\eta}[R(t)]\boldsymbol{\Psi}(t)=
    \boldsymbol{\Phi}(t_0)^\dagger\bG(t)^\dagger
    \boldsymbol{\eta}[R(t)]\bG(t)\boldsymbol{\Psi}(t_0).
    \label{eq21}
    \ee
If the connection $\cA$ is compatible with the Hermitian (metric) structure on $\cE$, $\br\phi(t),\psi(t)\kt_{R(t)}$ must not depend on $t$. Because this condition should hold for arbitrary choices of  $\phi(t_0),\psi(t_0)\in \cH_{R(t_0)}$,  it is equivalent to:
    \be
    \frac{d}{dt}[\bG(t)^\dagger\boldsymbol{\eta}[R(t)]\bG(t)]=\bzero,
    \label{eq22}
    \ee
where $\bzero$ stands for the $N\times N$ null matrix.
With the help of (\ref{eq7}) and the fact that (\ref{eq22}) holds for every (smooth) curve $\cC$, we can use it to arrive at the following metric-compatibility condition for the local connection one-forms $\bA$:
    \be
    \bA^\dagger-\boldsymbol{\eta}\bA\boldsymbol{\eta}^{-1}
    =i(d\boldsymbol{\eta})\boldsymbol{\eta}^{-1}.
    \label{eq23}
    \ee
where $d\boldsymbol{\eta}=\sum_{a=1}^d\partial_a\boldsymbol{\eta}\:dR^a$.

The following observations give the general form of metric-compatible local connection one-forms
$\bA$:
    \begin{enumerate}
    \item We can satisfy~(\ref{eq23}) by setting $\bA=\bA_0$ where
        \be
        \bA_0:=-\frac{i}{2}\,\boldsymbol{\eta}^{-1}d\boldsymbol{\eta}.
        \label{A-zero}
        \ee
    $\bA_0$ fulfils the $\boldsymbol{\eta}$-pseudo-anti-Hermiticity relation:
        \be
        \bA_0^\dagger=-\boldsymbol{\eta}\,\bA_0\boldsymbol{\eta}^{-1},
        \label{anti-ph}
        \ee
    and the identity:
        \be
        d\bA_0+2i\bA_0\wedge \bA_0=\bzero,
        \label{zero-curvature}
        \ee
    where $d$ and $\wedge$ are respectively the exterior derivative and wedge product of
    matrix-valued one-forms. Because the local curvature two-form $\bF_{\bA}$ of a connection
    one-form $\bA$ is given by \cite{cecile}:
        \be
        \bF_{\bA}=d\bA+i\bA\wedge\bA,
        \label{curvature}
        \ee
    (\ref{zero-curvature}) means that
        \[\bF_{\bA_0}=-i\bA_0\wedge\bA_0=\frac{1}{2}d\bA_0.\]
    \item Every local connection one-form satisfying (\ref{eq23}) has the form
        \be
        \bA=\bA_0+\boldsymbol{\omega}
        \label{gen-A}
        \ee
    where $\boldsymbol{\omega}$ is a matrix-valued one-form\footnote{Under a coordinate transformation $R\to\tilde R$, the components of $\boldsymbol{\omega}$ transform according to:
    $\boldsymbol{\omega}_a[R]\to\tilde{\boldsymbol{\omega}}_a[\tilde R]:=
    \sum_{b=1}^d\frac{\partial R^b}{\partial \tilde R^a}
    \boldsymbol{\omega}_b[R]\Big|_{R=R(\tilde R)}$.}
fulfiling the $\boldsymbol{\eta}$-pseudo-Hermiticity condition:
        \be
        \boldsymbol{\omega}_j^\dagger=\boldsymbol{\eta}\,\boldsymbol{\omega}_j
        \boldsymbol{\eta}^{-1}.
        \label{omega-ph}
        \ee
    \end{enumerate}
These observations show that every metric-compatible local connection one-form admits a unique decomposition  into the sum of an $\boldsymbol{\eta}$-pseudo-Hermitian and  an  $\boldsymbol{\eta}$-pseudo-anti-Hermitian matrix-valued one-form.

Next, we express (\ref{eq23}) in terms of certain associated linear operators acting in the typical fiber $V=\C^N$. To do this, first we choose the local basis sections $\psi_n[R]$ in such a way that the isomorphism $\varphi_{\alpha,R}:V_R\to\C^N$ defining the local trivialization $(\cO_\alpha,f_\alpha)$ maps $\psi_n[R]$ to the standard basis vectors $e_n$ of $\C^N$. In other words, we set
    \be
    \psi_n[R]:=\varphi_{\alpha,R}^{-1}(e_n).
    \label{psi-n=}
    \ee
This implies that for every $\psi_R\in V_R$,
    \[\varphi_{\alpha,R}(\psi_R)=\Psi[R]:=\sum_{n=1}^N\Psi_n[R]e_n=
    (\Psi_1[R],\Psi_2[R],\cdots,\Psi_n[R])\in\C^N,\]
where $\Psi_j[R]$'s are the components of $\psi_R$ in the basis $\{\psi_n[R]\}$. Next, we endow the typical fiber $\C^N$ with the Euclidean inner product (\ref{euclidean}) and label the resulting inner-product space by $\sH$. We can use $\varphi_{\alpha,R}$ to associate a linear operator $\eta[R]:\sH\to\sH$ to the matrix $\boldsymbol{\eta}[R]$. We identify it with the operator that maps each $z:=(z_1,z_2,\cdots,z_N)\in\C^N$ to
    \be
    \eta[R]\,z:=\sum_{m,n=1}^N\eta_{mn}[R] z_n e_m=\sum_{m,n=1}^N\br\psi_m[R],\psi_n[R]\kt_R\,z_n e_m
    =\sum_{m=1}^N\br\varphi_{\alpha,R}^{-1}(e_m), \varphi_{\alpha,R}^{-1}(z)\kt_R\: e_m.
    \label{eta=}
    \ee
In view of (\ref{eta-mn}), $\eta[R]$  is a positive-definite (metric) operator \cite{review} acting in $\sH$. Therefore it defines the following inner product on the typical fiber $\C^N$.
    \be
    \br\cdot|\cdot\kt_{\eta[R]}:=\br\cdot|\eta[R]\cdot\kt.
    \label{eq24}
    \ee

Similarly, we use the entries of $\bA_a[R]$ to introduce the linear operators $A_a[R]:\sH\to\sH$ given by
    \be
     A_a[R]\,z:=\sum_{m,n=1}^N A_{amn}[R] z_n e_m.
     \label{A-sub-a}
     \ee
We can assemble these to define the operator-valued one-form
    \be
    A=  \sum_{a=1}^dA_a dR^a,
    \label{one-form-A}
    \ee
and use $\varphi_{\alpha,R}$ to express the compatibility relation (\ref{eq23}) in terms of $\eta$ and $A$ as
    \be
    A^\dagger- \eta A \eta^{-1}=id\eta\: \eta^{-1},
    \label{eq25}
    \ee
where $A^\dagger:= \sum_{a=1}^d A_a^\dagger dR^a$, the superscript $\dagger$ on an operator acting in $\sH$ stands for is adjoint, and $d\eta:= \sum_{a=1}^d\partial_a\eta\: dR^a$. Again the general solution of (\ref{eq25}) has the form
    \be
    A=A_0+\omega,
    \label{gen-A-2}
    \ee
where $A_0:=-(i/2) {\eta}^{-1}d{\eta}$, which is an $\eta$-pseudo-anti-Hermitian operator-valued one-form, i.e., it fulfills:
    \be
    A_0^\dagger=-\eta\,A_0\eta^{-1},
    \label{anti-ph-2}
    \ee
while $\omega$ is an $\eta$-pseudo-Hermitian operator-valued one-form, i.e.,
    \be
    \omega^\dagger=\eta\,\omega\,\eta^{-1}.
    \label{Azero-ph-2}
    \ee

Note that if $\sH_{\eta[R]}$ is the inner-product space obtained by endowing $\C^N$ with the inner product (\ref{eq24}), then the isomorphism $\varphi_{\alpha,R}$ viewed as an operator mapping $\cH_R$ onto $\sH_{\eta[R]}$ 
is unitary. We can use this operator to express Eq.~(\ref{con-der}) for the parallel transportation as the Schr\"odinger equation:
    \be
    i\frac{d}{dt}\Psi(t)=H_{A}(t)\Psi(t),
    \label{sch-eq-A}
    \ee
defined by the Hamiltonian operator
    \be
    H_A(t):=\sum_{a=1}^d \dot R^a(t)A_a[R(t)]
    \label{H-A}
    \ee
in the Hilbert space $\sH_{\eta[R]}$. In view of (\ref{eq25}) and (\ref{H-A}),
    \be
    H_A^\dagger- \eta H_A \eta^{-1}=i\dot\eta\, \eta^{-1},
    \label{eq31}
    \ee
where $\dot\eta:=\sum_{a=1}^d\dot R^a\partial_a\eta$.

The resemblance of (\ref{eq2}) and (\ref{eq31}) is to be expected, for both are the result of the conservation of the inner product of vectors. Yet there is a major difference between $H$ and $H_A$, namely that the Schr\"odinger equation (\ref{sch-eq-A}) determined by $H_A$ is invariant under reparameterizations of $t$, i.e., it generates a purely geometric evolution that is sensitive only to the local connection one-form $\bA$ and the curve $\cC$. In particular, it does not depend on the time it takes to traverse the segment of the curve $\cC$ lying between $R(t_0)$ and $R(t)$.

The geometric evolution generated by $H_A$ in $\sH_{\eta[R(t)]}$ corresponds to the parallel transportation of $\psi(t_0)$ in the vector bundle $\cE$. This defines a curve in $\cE$ that is called the horizontal lift of $\cC$, \cite{cecile}. In a sense $H_A$ is the generator of the ``horizontal evolution'' of the states of the system.

Now, consider the case that the state space of our system seizes to be time-dependent. This corresponds to fixing a single point $R_0$ in $M$ and confining our attention to the fiber $V_{R_0}$. In this case, we do not need to use a connection to define the rate of change of the state vectors, because their dynamics occurs in $V_{R_0}$. In this sense, the Hamiltonian operator that defines the dynamics of our system generates ``vertical evolution'' of the states.

Our observations regarding the horizontal and vertical evolutions lead us to postulate that {\em the dynamics of a general quantum system with a time-dependent state space involves both horizontal and vertical evolutions}. More specifically, we propose to take the Hamiltonian operator $H(t)$, which generates the time-evolution of the system through the Schr\"odinger equation (\ref{sch-eq}), in the form
    \be
    H(t)=H_A(t)+H_E(t),
    \label{H=HH}
    \ee
where $H_A(t)$ is a geometric Hamiltonian, that is determined by a metric-compatible connection via (\ref{H-A}), and $H_E(t)$ is an operator that describes the interaction of the system with the forces that are not related to the time-dependence of its state space. We call the latter: ``direct interactions.'' They are responsible for the vertical evolutions and unlike the indirect interactions quantified by $H_A(t)$ survive when the state space of the system becomes time-independent.

An immediate consequence of (\ref{eq2}), (\ref{eq31}), and (\ref{H=HH}) is that, unlike $H$ and $H_A$, the operator $H_E$ satisfies the $\eta$-pseudo-Hermiticity relation:
    \be
    H_E^\dagger=\eta H_E\eta^{-1}.
    \ee
This means that as an operator acting in $\sH_{\eta}$, $H_E$ is Hermitian. We therefore identify it with the {\em energy operator} of our system. In view of (\ref{H=HH}), this implies that to determine the energy of a quantum system with a time-dependent state space we need both the Hamiltonian of the system $H$, which generates its dynamics, and the metric-compatible connection which specifies $H_A$.

According to (\ref{gen-A-2}), the geometric Hamiltonian $H_A$ has the general form
    \be
    H_A(t)=H_{A_0}(t)+H_\omega(t),
    \label{eq41}
    \ee
where
    \begin{align}
    &H_{A_0}(t):=
    -\frac{i}{2}{\eta (t)}^{-1}\dot{\eta}(t),
    &&H_\omega(t):=\sum_{a=1}^d\dot R^a(t)\omega_a(t),
    \label{eq42}
    \end{align}
and $\omega_a(t):=\omega_a[R(t)]$. Substituting (\ref{eq41}) in (\ref{H=HH}) gives
the structure of the Hamiltonian of the system:
    \be
    H(t)=H_E(t)+H_\omega(t)+H_{A_0}(t)=H_{ph}(t)+H_{A_0},
    \label{H=}
    \ee
where $H_{ph}(t)$ denotes the $\eta$-pseudo-Hermitian part of $H(t)$, i.e.,
    \be
    H_{ph}(t):=H_E(t)+H_\omega(t).
    \label{H-ph=}
    \ee

The above analysis shows that the Hilbert space-Hamiltonian operator pair $(\sH_{\eta(t)},H(t))$ provides a local description of our quantum system in which its state space is time-dependent. Following the prescription outlined in \cite{jpa-2003} we can also obtain a local Hermitian representation of our system by performing a unitary transformation to map $\sH_{\eta(t)}$ onto $\sH$.

It is easy to check that $\rho(t):=\sqrt{\eta(t)}:\sH_{\eta(t)}\to\sH$ is a unitary operator and that it maps the solutions $\Psi(t)$ of the Schr\"odinger equation~(\ref{sch-eq-H-last}) onto the solutions of the Schr\"odinger equation
    \be
    i\frac{d}{dt}\Phi(t)=h(t)\Phi(t),
    \label{sch-eq-h-last}
    \ee
for the Hamiltonian operator
    \be
    h(t):=h_A(t)+h_E(t),
    \label{h=}
    \ee
where
    \bea
    h_A(t)&:=&\rho(t)H_A(t)\rho^{-1}+i\dot\rho(t)\rho(t)^{-1}
    =\rho(t)H_\omega(t)\rho^{-1}+\frac{i}{2}[\dot\rho(t),\rho(t)^{-1}],
    \label{little-h-A}\\
    h_E(t)&:=&\rho(t)H_E(t)\rho^{-1}.
        \label{little-h-E}
        \eea
Inserting these equations in (\ref{h=}) and making use of (\ref{H-ph=}), we have
    \be
    h(t):=\rho(t)H_{ph}(t)\rho^{-1}+\frac{i}{2}[\dot\rho(t),\rho(t)^{-1}].
    \label{h=2}
        \ee
Because $H_{ph}(t):\sH\to\sH$ is $\eta(t)$-pseudo-Hermitian, the first term on the right-hand side of this relation acts as a Hermitian operator in $\sH$. The same holds for the second term, because $\rho(t):\sH\to\sH$ and consequently $\dot\rho(t):\sH\to\sH$ are Hermitian operators. This proves the Hermiticity of $h(t):\sH\to\sH$ and the unitarity of the Schr\"odinger time-evolutions it generates in $\sH$.

It is important to notice that the unitary operator $\rho(t):\sH_{\eta(t)}\to\sH$ does not map the energy operator $H_E(t)$ to $h(t)$. It maps it to the non-geometric part of $h(t)$ that is given by (\ref{little-h-E}). In general $h(t)$ also includes a geometric part, namely $h_A(t)$, that is determined by the metric on $\cE$, a corresponding compatible connection $\cA$, and the segment of the curve $\cC$ in $M$ that lies in $\cO_\alpha$. Because $h_E(t)$ and $h_A(t)$ act as Hermitian operators in $\sH$, they represent observables of the system.

\section{Geometric extension of quantum mechanics}
\label{S3}

The above description of quantum dynamics is clearly local in the sense that we have confined our analysis to the segment of the curve $\cC$ that lies in a single local coordinate patch. To consider more general situations where $\cC$ lies in the union of different local coordinate patches, we must use the transition functions of $\cE$ to glue together the pieces of the bundle that lie above the intersection of these patches. This allows for defining the dynamics globally provided that, in addition to a globally defined connection on $\cE$, we have a global prescription for determining the energy observable. As we explain below, we can achieve this by assigning a Hermitian operator $\fh_E[R]:\cH_R\to\cH_R$ to each $R\in M$. This corresponds to a global section of a real vector bundle over $M$ whose fibers are the real vector space of Hermitian operators acting in fibers of $\cE$. In order to give a more detailed description of  this vector bundle, which we denote by $\fu(\cE)$, we need some preparation.

Let us consider an arbitrary pair of intersecting local patches of $M$, say $\cO_\alpha$ and $\cO_\beta$, and the corresponding local trivialization maps $\varphi_{\alpha,R}:\cH_R\to V$ and $\varphi_{\beta,R}:\cH_R\to V$. Using $\varphi_{\beta,R}$ in place of $\varphi_{\alpha,R}$ in (\ref{eta=}), we obtain a positive-definite operator $\tilde\eta[R]:\sH\to\sH$ satisfying
    \be
        \tilde\eta[R]\,z:=\sum_{m=1}^N\br\varphi_{\beta,R}^{-1}(e_m), \varphi_{\beta,R}^{-1}(z)\kt_R\: e_m.
    \label{t-eta=}
    \ee
This operator determines an inner product on $V$, namely $\br\cdot,\cdot\kt_{\tilde\eta[R]}:=\br\cdot|\tilde\eta[R]\cdot\kt$. Endowing $V$ with this inner product gives an inner-product space that we label by $\sH_{\tilde\eta[R]}$.
For $R\in\cO_\alpha\cap\cO_\beta$, we can express $\tilde\eta[R]$ in terms of $\eta[R]$ and the transition function $g_{\alpha\beta,R}$. To achieve this, first we use (\ref{t-eta=}) and the unitarity of $\varphi_{\alpha,R}:\cH_R\to \sH_{\eta[R]}$ to show that
    \bea
    \br e_m|\, \tilde\eta[R]\, e_n\kt&=&
    \br\varphi_{\beta,R}^{-1} \, e_m\,, \varphi_{\beta,R}^{-1}\,e_n\,\kt_R=
    \br\varphi_{\alpha,R}\,\varphi_{\beta,R}^{-1}\,e_m\,,\varphi_{\alpha,R}
    \, \varphi_{\beta,R}^{-1}\,e_n\,\kt_{\eta[R]}\nn\\
    &=&\br g_{\alpha\beta,R}\, e_m,g_{\alpha\beta,R} \, e_n\,\kt_{\eta[R]}
    =\br g_{\alpha\beta,R}\, e_m|\,\eta[R]\,g_{\alpha\beta,R}\, e_n\,\kt\nn\\
    &=&\br\, e_m\, |\, g_{\alpha\beta,R}^\dagger\, \eta[R]\,g_{\alpha\beta,R}\,e_n\,\kt.
    \nn
    \eea
Accordingly, the matrix representation of $\tilde\eta[R]$, $g_{\alpha\beta,R}$, and $\eta[R]$ in the basis $\{e_n\}$, which we respectively label by $\tilde{\boldsymbol{\eta}}[R]$, $\bg_{\alpha\beta,R}$, and $\boldsymbol{\eta}[R]$, fulfil
    \be
    \tilde{\boldsymbol{\eta}}[R]=\bg_{\alpha\beta,R}^\dagger\,\boldsymbol{\eta}[R]\,\bg_{\alpha\beta,R}.
    \label{matrix-rep-eta-g-eta}
    \ee
This implies that  $\tilde\eta[R]=g_{\alpha\beta,R}^\dagger\, \eta[R]\,g_{\alpha\beta,R}$, where $g_{\alpha\beta,R}^\dagger$ denotes the adjoint of the operator $g_{\alpha\beta,R}$ viewed as mapping $\sH$ to $\sH$, i.e.,
$g_{\alpha\beta,R}^\dagger:\sH\to\sH$ is the linear operator satisfying $\br w|g_{\alpha\beta,R}^\dagger\,z\kt=\br g_{\alpha\beta,R}\:w|z\kt$ for all $w,z\in\sH$.\footnote{Notice that the definition of the adjoint of an operator depends on the inner product on its domain and range.}

Similarly to $\varphi_{\alpha,R}:\cH_R\to \sH_{\eta[R]}$, the isomorphism $\varphi_{\beta,R}:\cH_R\to \sH_{\tilde\eta[R]}$ is a unitary operator. This in turn implies that the transition function $g_{\alpha\beta,R}=\varphi_{\alpha,R}\,\varphi_{\beta,R}^{-1}$ viewed as mapping $\sH_{\tilde\eta[R]}$ onto $\sH_{\eta[R]}$ is a unitary operator. Moreover, the positive square root of $\eta[R]$ and $\tilde\eta[R]$, that we denote by $\rho[R]$ and $\tilde\rho[R]$,  define unitary operators mapping $\sH_{\eta[R]}$ to $\sH$ and $\sH_{\tilde\eta[R]}$ to $\sH$, respectively. Because
    \begin{align*}
    &g_{\alpha\beta,R}:\sH_{\tilde\eta[R]}\to\sH_{\eta[R]},
    &&\rho[R]:=\sqrt{\eta[R]}:\sH_{\eta[R]}\to\sH,
    && \tilde\rho[R]:=\sqrt{\tilde\eta[R]}:\sH_{\tilde\eta[R]}\to\sH,
    \end{align*}
are unitary operators, so is the operator $\cG_{\alpha\beta,R}:\sH\to\sH$ defined by
        \be
        \cG_{\alpha\beta,R}:=
        \rho[R]\: g_{\alpha\beta,R}\:\tilde\rho[R]^{-1}
        .
        \label{cG-unitary}
        \ee

We define $\fu(\cE)$ as the real vector bundle over $M$ whose fiber over $R\in M$ is the real vector space $u(\cH_R)$ of Hermitian operators acting in $\cH_R$.
The typical fiber of $\fu(\cE)$, which we denote by $u(\sH)$, is the $N^2$-dimensional real vector space of Hermitian operators acting in $\sH$.\footnote{As a real vector space, this is isomorphic to the Lie algebra $u(n)$.} The transition functions of $\fu(\cE)$ are isomorphisms $\fg_{\alpha\beta,R}: u(\sH)\to u(\sH)$ defined by
    \be
    \fg_{\alpha\beta,R}(\fH):=\cG_{\alpha\beta,R}\, \fH \, \cG_{\alpha\beta,R}^{-1},
    \label{transition-u}
    \ee
where $\cG_{\alpha\beta,R}:\sH\to\sH$ is given by (\ref{cG-unitary}), and $\fH$ is an arbitrary element of $u(\sH)$. Note that because $\cG_{\alpha\beta,R}:\sH\to\sH$ is a unitary operator, (\ref{transition-u}) defines a linear operator mapping $u(\sH)$ onto $u(\sH)$. It is also easy to see that this operator is a vector space isomorphism.

The above constructions suggest an extension of quantum mechanics in which the role of the Hilbert space and the Hamiltonian operator is respectively played by the Hermitian vector bundle $\cE$, which is endowed with a metric-compatible connection $\cA$, and a global section of the real vector bundle $\fu(\cE)$. In the following we outline its basic postulates.
    \begin{enumerate}
    \item A quantum system $\cS$ is described by a triplet $(\cE,\cC,\fh_E)$, where $\cE$ is a Hermitian vector bundle $\cE$ endowed with a metric-compatible connection $\cA$, $\cC:[t_1,t_2]\to M$ is a smooth curve in the base manifold $M$ of $\cE$, $[t_1,t_2]$ is the time interval during which we wish to describe the system, and $\fh_E:M\to\fu(\cE)$ is a global section of the real vector bundle $\fu(\cE)$. As we have explained above, the latter is uniquely determined by $\cE$.
    \item $\cE$ determines the kinematic properties of the system. Let $t_\star\in[t_1,t_2]$ be an arbitrary instant of time. Then the pure states of the quantum system at time $t_\star$ are rays (one-dimensional subspaces) of the Hilbert space $\cH_{R_\star}:=\cH_{\cC(t_\star)}$, i.e., the fiber of $\cE$ over $R_\star:=\cC(t_\star)\in M$. The observables of the system are the global sections of $\fu(\cE)$. To describe the measurement of an observable $\fO: M\to\fu(\cE)$ at time $t_\star$ in a pure state given by some $\psi_\star\in\cH_{R_\star}$, one evaluates $\fO$ at $R_\star$ and obtains a Hermitian operator $\fO_{R_\star}:=\fO[R_\star]$ acting in $\cH_{R_\star}$. One then makes use of von Neumann's projection axiom to predict the outcome of the measurement, i.e.,  compute the probability of obtaining its possible values, which are the eigenvalues of $\fO_{R_\star}$, as well as their average (expectation value of the observable) and standard deviation (uncertainty in measuring the observable).\footnote{Extending the algebraic definition of a mixed state in standard quantum mechanics \cite{faddeev}, we identify the mixed states with the global sections $\Omega:M\to u(\cH)^*$ of the dual vector bundle \cite{dual-VB} to $u(\cH)$ such that for every $R\in M$ and $O\in u(\cH_{R})$ the linear functional $\Omega_R:=\Omega[R]:u(\cH_{R})\to\R$
satisfies $\Omega_R(I_R)=1$ and $\Omega_R(O^2)\geq 0$, where $I_R$ is the identity operator acting in $\cH_{R}$. The expectation value of an observable $\fO$ in a mixed state $\Omega$ at time $t_\star$ is given by $\Omega_{R_\star}(\fO_{R_\star})$.}
    \item The dynamics of the system is described by a lift $\psi:[t_1,t_2]\to\cE$ of the curve $\cC:[t_1,t_2]\to M$ to $\cE$ that is determined by the covariant Schr\"odinger equation:
        \be
        i\hbar D_t\psi(t)=\fh_E[\cC(t)]\psi(t).
        \label{cov-sch-eq}
        \ee
Here the term `lift' means that $\psi(t)\in\cH_{\cC(t)}$, and $D_t$ is the covariant time-derivative determined by the connection $\cA$.
        \end{enumerate}

For time intervals where $\cC(t)$ lies within a single local coordinate patch $\cO_\alpha$, we can choose the basis consisting of the local sections $\psi_n:\cO_\alpha\to\cE$ and express $\psi(t)$ in terms of its components $\Psi_n(t)$ in this basis according to
    \be
    \psi(t)=\sum_{n=1}^N \Psi_n(t)\psi_n[R(t)],
    \label{local-psi}
    \ee
where we identify $\cC(t)$ with its coordinates $R(t)$ in $\cO_\alpha$. Substituting (\ref{local-psi}) in (\ref{cov-sch-eq}), we find
    \be
    i\bD_t\boldsymbol{\Psi}(t)=\bH_E[R(t)]\boldsymbol{\Psi}(t),
    \label{local-sch-eq-123}
    \ee
where $\bD_t$ is defined in (\ref{cov-der=}) and $\bH_E[R]$ is the $N\times N$ matrix with entries
    \be
    H_E[R]_{mn}:=\br\psi_m[R],\fh_E[R]\psi_n[R]\kt_{R}.
    \label{HE-d=}
    \ee

Let us consider the linear operator $H_E[R]:\sH\to\sH$ that is defined by
    \be
    H_E[R]:=\varphi_{\alpha,R}\;\fh_E[R]\;\varphi_{\alpha,R}^{-1}.
    \label{HE-d=}
    \ee
It is not difficult to show that for all $z=\sum_{n=1}^Nz_ne_n\in\C^N$, $H_E[R]z=\sum_{m,n=1}^N H_E[R]_{mn}z_ne_m$.
We can view $H_E(t)$ as a linear operator acting in $\sH_{\eta(t)}:=\sH_{\eta[R(t)]}$. Because $\varphi_{\alpha,R}:\cH_R\to\sH_{\eta[R]}$ and $\fh_E[R]:\cH_R\to \cH_R$  are respectively unitary and Hermitian operators,  Eq.~(\ref{HE-d=}) implies that $H_E(t):\sH_{\eta(t)}\to\sH_{\eta(t)}$ is a Hermitian operator.

Next, we recall that the connection $\cA$ on $\cE$ determines a local connection one-form $\bA$ in the patch $\cO_\alpha$. Using the entries of $\bA$ in (\ref{A-sub-a}), we find the operators $A_a[R]$. Substituting these in (\ref{H-A}), we obtain $H_A(t)$ which acts in $\sH_{\eta(t)}:=\sH_{\eta[R(t)]}$ as a Hermitian operator. We can use the argument establishing the equivalence of (\ref{con-der}) and (\ref{sch-eq-A}) to express (\ref{local-sch-eq-123}) in the form
    \be
    i\frac{d}{dt}\Psi(t)=H(t)\Psi(t),
    \label{sch-eq-H-last}
    \ee
where $H(t):=H_A(t)+H_E(t)$, $H_E(t):=H_E[R(t)]$, and $H_E[R]$ is given by (\ref{HE-d=}).

As we discussed in Sec.~\ref{S2}, $H_A(t)$ and $H_E(t)$ respectively give the geometric part of the Hamiltonian and the energy operator of the system at time $t$ in the local patch $\cO_\alpha$. Our analysis shows that these are respectively determined by the connection $\cA$ on $\cE$ and the global section $\fh_E$ of $\fu(\cE)$. Given $\cA$ and $\fh_E$ we can describe our quantum system using the Hilbert space-Hamiltonian operator pair $(\sH_{\eta(t)},H(t))$. This is a local description that is valid only for the segment of the curve $\cC$ that lines in the coordinate patch $\cO_\alpha$. We can also obtain an equivalent local description using the Hilbert space $\sH$ and the Hermitian operator $h(t):\sH\to\sH$ given by (\ref{h=}). This is the standard textbook description of a quantum system. Note however that similarly to the description provided by $(\sH_{\eta(t)},H(t))$, it applies locally, i.e., it involves the use of a single local trivialization of $\cE$. If the curve $\cC(t)$ lies in the union of two or more patches, we must obtain the analog of the Hermitian Hamiltonian $h(t)$ in all of these patches and try to relate them in their intersection. This requires the knowledge of the transformation rule for the operators $H(t)$ and $h(t)$ under changes of local trivializations.\footnote{Alternatively, one might be able to use a different covering of $M$ by coordinate patches and a corresponding set of local trivializations of $\cE$ such that the curve $\cC$ is contained in a single coordinate patch.}

Consider a pair of intersecting local patches and the corresponding local trivializations, $(\cO_\alpha,f_\alpha)$  and $(\cO_\beta,f_\beta)$. For each $R\in\cO_\alpha\cap\cO_\beta$ and $\psi_R\in\cH_R$,
we have $\Psi:=\varphi_{\alpha,R}(\psi_R)\in\sH$. This is the representation of $\psi_R$ in the local trivialization $(\cO_\alpha,f_\alpha)$. Similarly, we have $\tilde\Psi:=\varphi_{\beta,R}(\psi_R)\in\sH$ as the representation of $\psi_R$ in the local trivialization $(\cO_\beta,f_\beta)$. It is easy to show that
    \be
    \tilde\Psi=g_{\beta\alpha,R}\Psi=g_{\alpha\beta,R}^{-1}\Psi.
    \label{transform}
    \ee
Let us recall that the transition function $g_{\alpha\beta,R}$ viewed as mapping $\sH_{\tilde\eta[R]}$ onto  $\sH_{\eta[R]}$  is a unitary operator. This shows that if we identify $\Psi$ and $\tilde\Psi$ respectively with elements of  $\sH_{\eta[R]}$  and $\sH_{\tilde\eta[R]}$, the transformation of local representation of the state vector $\psi[R]$, namely $\Psi\to\tilde\Psi$, corresponds to a unitary transformation.

In view of (\ref{connection-trans}), we can express the operator-valued one-form $\tilde A[R]$ associated with the local connection one-form $\tilde\bA$ as
    \be
    \tilde A[R]=g_{\alpha\beta,R}^{-1}\: A[R] \: g_{\alpha\beta,R}
    -ig_{\alpha\beta,R}^{-1}\: dg_{\alpha\beta,R},
    \label{tilde-A}
    \ee
where $dg_{\alpha\beta,R}:=\sum_{a=1}^d\partial_a g_{\alpha\beta,R}dR^a$. Eq.~(\ref{tilde-A}) implies that the operator defined by $\tilde H_{\tilde A}(t):= \dot R^j(t)\tilde A_j[R(t)]$, which is the analog of $H_A(t)$ for the local trivialization $(\cO_\beta,f_\beta)$, satisfies
    \be
    \tilde H_{\tilde A}(t)= g_{\alpha\beta,R(t)}^{-1}\: H_A(t) \: g_{\alpha\beta,R(t)}
    -ig_{\alpha\beta,R(t)}^{-1}\: \dot g_{\alpha\beta,R(t)},
    \label{tilde-H-A}
    \ee
where $\dot g_{\alpha\beta,R(t)}:=\sum_{a=1}^d\dot R^a(t)\partial_a g_{\alpha\beta,R}\big|_{R=R(t)}$.

In the local trivialization $(\cO_\beta,f_\beta)$ the energy observable is represented by an operator $\tilde H_E(t):\sH_{\tilde\eta(t)}\to\sH_{\tilde\eta(t)}$ that we can determine using $\varphi_{\beta,R}$ in place of $\varphi_{\alpha,R}$ in (\ref{HE-d=}). This gives
    \be
    \tilde H_E(t)=\varphi_{\beta,R}\;\fh_E[R]\;\varphi_{\beta,R}^{-1}=
    g_{\alpha\beta,R(t)}^{-1}\: H_E(t)\:g_{\alpha\beta,R(t)}.
    \label{tilde-H-E}
    \ee
Equations (\ref{tilde-H-A}) and (\ref{tilde-H-E}) show that the operator defined by
    \be
    \tilde H(t):=\tilde H_{\tilde A}(t)+\tilde H_E(t)
    \label{tilde-H-def}
    \ee
fulfills
    \be
    \tilde H(t)=g_{\alpha\beta,R(t)}^{-1}\: H(t) \: g_{\alpha\beta,R(t)}
    -ig_{\alpha\beta,R(t)}^{-1}\: \dot g_{\alpha\beta,R(t)}.
    \label{tilde-H}
    \ee

Now, consider the local representation of the evolving state vector provided by the local trivialization $(\cO_\beta,f_\beta)$, i.e., $\tilde\Psi(t)=g_{\alpha\beta,R(t)}^{-1}\Psi(t)$. Equation (\ref{tilde-H}) ensures that
    \be
    i\frac{d}{dt}\tilde\Psi(t)=\tilde H(t)\tilde\Psi(t)
    \label{sch-eq-H-last-tilde}
    \ee
if and only if $\Psi(t)$ is a solution of (\ref{sch-eq-H-last}). This demonstrates the consistency of the formulation of the dynamics using the covariant Schr\"odinger equation  (\ref{cov-sch-eq}). In particular, in each local patch we have a representation of the quantum system that is given by $(\sH_{\eta(t)},H(t))$ and when $\cC(t)$ belongs to the intersection of two or more local patches we can represent the system by the local representation associated with any of them in a consistent manner.

Next, consider a curve $\cC$ such that $\cC(t)\in\cO_\alpha\cup\cO_\beta$, $\cC(t_1)\in\cO_\alpha\setminus\cO_\beta$, and $\cC(t_2)\in\cO_\beta\setminus\cO_\alpha$. Suppose that at $t=t_1$ the system is in a state that is locally represented by $\Psi_1\in\sH_{\eta(t_1)}$. To determine the evolution of the system we proceed as follows.
    \begin{itemize}
    \item[] i) Start from the representation  $(\sH_{\eta(t)},H(t))$, solve the Schr\"odinger equation
(\ref{sch-eq-H-last}) together with the initial condition $\Psi(t_1)=\Psi_1$ to obtain $\Psi(\tau)$ for some $\tau$ such that $\cC(\tau)\in\cO_\alpha\cap\cO_\beta$.
    \item[]ii)  Switch to the representation  $(\sH_{\tilde\eta(t)},\tilde H(t))$ at $t=\tau$, take
$\tilde\Psi(\tau):=g_{\alpha\beta,R(\tau)}^{-1}\Psi(\tau)$ as the initial condition for (\ref{sch-eq-H-last-tilde}), and solve this equation for $t>\tau$ to determine $\tilde\Psi(t_2)$.
    \end{itemize}
Let us examine the description of the evolution using the Hermitian representations $(\sH,h(t))$ and $(\sH,\tilde h(t))$ associated with the $(\sH_{\eta(t)},H(t))$ and $(\sH_{\tilde\eta(t)},\tilde H(t))$.

The operators $h(t):\sH\to\sH$ and $\tilde h:\sH\to\sH$ are respectively given by
    \begin{align}
    & h(t)=\rho(t)H(t)\rho(t)^{-1}+i\dot\rho(t)\rho(t)^{-1},
    \label{h=11}\\
    &\tilde h(t)=\tilde\rho(t)\tilde H(t)\tilde\rho(t)^{-1}+i\dot{\tilde\rho}(t)\tilde\rho(t)^{-1}.
    \label{th=11}
    \end{align}
Substituting  (\ref{tilde-H}) in (\ref{th=11}) and making use of (\ref{cG-unitary}) and (\ref{h=11}), we obtain
    \be
    \tilde h(t)=\cG_{\alpha\beta,R(t)}^{-1}h(t)\cG_{\alpha\beta,R(t)}-i
    \cG_{\alpha\beta,R(t)}^{-1}\dot\cG_{\alpha\beta,R(t)}.
    \label{tilde-h=h}
    \ee
In view of (\ref{h=}), we can write $\tilde h(t)=\tilde h_A(t)+\tilde h_E(t)$, where
    \bea
    \tilde h_A(t)&=&\cG_{\alpha\beta,R(t)}^{-1}h_A(t)\cG_{\alpha\beta,R(t)}-i
    \cG_{\alpha\beta,R(t)}^{-1}\dot\cG_{\alpha\beta,R(t)}
    \eea
and
    \bea
    \tilde h_E(t)&=&\cG_{\alpha\beta,R(t)}^{-1}h_E(t)\cG_{\alpha\beta,R(t)}=
    \fg_{\alpha\beta,R(t)}^{-1}(h_E(t))
    \label{th=ghg}
    \eea
are respectively the geometric part of $\tilde h(t)$ and the energy operator in the representation $(\sH,\tilde h(t))$, and $\fg_{\alpha\beta,R(t)}$ is the transition function (\ref{transition-u}) of the bundle $\fu(\cE)$.

The description of dynamics of the system using its Hermitian local representations $(\sH,h(t))$ and $(\sH,\tilde h(t))$ involves the following analog of the steps (i) and (ii).
    \begin{itemize}
    \item[] i$'$) Start from the representation  $(\sH,h(t))$, solve the Schr\"odinger equation
(\ref{sch-eq-h-last}) together with the initial condition $\Phi(t_1)=\Phi_1:=\rho(t_1)\Psi_1$ to obtain $\Phi(\tau)$ for some $\tau$ such that $\cC(\tau)\in\cO_\alpha\cap\cO_\beta$.
    \item[]ii$'$)  Switch to the representation  $(\sH,\tilde h(t))$ at $t=\tau$ using (\ref{tilde-h=h}), take $\tilde\Phi(\tau):=\cG_{\alpha\beta,R(\tau)}^{-1}\Phi(\tau)$ as the initial condition for the Schr\"odinger equation $i\dot{\tilde\Phi}(t)=\tilde h(t)\tilde\Phi(t)$, and solve this equation for $t>\tau$ to determine $\tilde\Phi(t_2)$. This is necessarily equal to $\tilde\rho(t)\tilde\Psi(t_2)$.
    \end{itemize}
By construction, the choice of $\tau$ does not change the outcome of this calculation. Notice however that for every generic choice of $\tau$, the evolution of the state vector $\Phi_1$ to $\Phi_2:=\tilde\Phi(t_2)$ defines a curve in $\sH$ that has a discontinuity at $t=\tau$, while the curve $\cC(t)$ is continuous.


An important aspect of the description of the quantum system using the local Hermitian representations $(\sH, h(t))$ and
$(\sH,\tilde h(t))$ is that if an observable of the system is quantified by the Hermitian operators $o[R]:\sH\to\sH$ and $\tilde o[R]:\sH\to\sH$ in the representations $(\sH, h(t))$ and $(\sH,\tilde h(t))$, then for each $R\in\cO_\alpha\cap\cO_\beta$, $o[R]$ and $\tilde o[R]$ are related via
    \be
    \tilde o(t)=\fg_{\alpha\beta,R}^{-1}(o[R])=\cG_{\alpha\beta,R}^{-1}\,o[R]\,\cG_{\alpha\beta,R}.
    \label{trans-observer}
    \ee
This means that although all the local Hermitian representations make use of the same Hilbert space $\sH$, the Hermitian operator representing an observable of the system in a local Hermitian representation depends on the choice of this representation.

\section{Implementation for two-level systems}
\label{S4}

In the geometric framework we have outlined above, a two-level quantum system is determined by a Hermitian vector bundle $\cE$ with typical fiber $V=\C^2$, a metric-compatible connection $\cA$ on $\cE$, and a global section $\fh_E$ of the real vector bundle $\fu(\cE)$. In this section we construct these mathematical structures and examine the dynamics of the system using its local Hermitian representations.

\subsection{Construction of $\cE$}

Because $V=\C^2$, for each $R\in M$ we can represent the metric operator $\eta[R]:\sH\to\sH$ by its matrix representation $\boldsymbol{\eta}[R]$ in the standard basis $\{e_1,e_2\}$. This is an $R$-dependent $2\times 2$ positive-definite matrix. We can express it in the form
    \be
    \boldsymbol{\eta}[R]=\bU[R]\,\boldsymbol{\eta}_d[R]\bU[R]^\dagger,
    \label{matrix-eta=}
    \ee
where $\bU[R]$ and $\boldsymbol{\eta}_d[R]$ are respectively unitary and diagonal $2\times 2$ matrices. The latter has the general form
    \be
    \boldsymbol{\eta}_d[R]=\left[\begin{array}{cc}
    \xi^2 & 0 \\
    0 & \zeta^2\end{array}\right]=\chi_+\bI+\chi_-\bsigma_3,
    \label{diag-eta=}
    \ee
where $\xi$ and $\zeta$ are positive real numbers that in general depend on $R$, $\bI$ is the $2\times 2$ identity matrix,
    \begin{align*}
    &\bsigma_1:=\left[\begin{array}{cc}
    0 & 1\\
    1 & 0\end{array}\right],
    &&\bsigma_2:=\left[\begin{array}{cc}
    0 & -i\\
    i & 0\end{array}\right],
    &&\bsigma_3:=\left[\begin{array}{cc}
    1 & 0\\
    0 & -1\end{array}\right],
    \end{align*}
are Pauli matrices, and
    \be
    \chi_\pm:=\frac{1}{2}(\xi^2\pm\zeta^2).
    \label{chi=}
    \ee
The unitary matrix $\bU[R]$ is not unique. Following \cite{book2}, we identify it with
    \be
    \bcU(\vartheta,\varphi):=e^{-i\varphi\bsigma_3/2} e^{-i\vartheta\bsigma_2/2}
    e^{i\varphi\bsigma_3/2}=
    \left[\begin{array}{cc}
    \cos(\vartheta/2)&-e^{-i\varphi}\sin(\vartheta/2)\\
    e^{i\varphi}\sin(\vartheta/2) & \cos(\vartheta/2)\end{array}\right],
    \label{bcU=}
    \ee
where $\vartheta$ and $\varphi$ are respectively polar and azimuthal angles in the spherical coordinates, i.e., we set
    \be
    \bU[R]:=\bcU(\vartheta,\varphi).
    \label{U=}
    \ee
In view of (\ref{bcU=}), this implies
    \be
    \bU[R]^\dagger=\bcU(-\vartheta,\varphi).
    \label{U-dagger=}
    \ee

Next, we introduce
    \be
    \bs_j(\vartheta,\varphi):=
    \bcU(\vartheta,\varphi)\,\bsigma_j\,\bcU(-\vartheta,\varphi)
    =\bU[R]\,\bsigma_j\,\bU[R]^\dagger,
    \label{bs=}
    \ee
and use (\ref{bcU=}) -- (\ref{U-dagger=}) and the identity:
    \be
    e^{-i\theta\bsigma_i/2}\bsigma_j\,e^{i\theta\bsigma_i/2}=\left\{\begin{array}{ccc}
    \bsigma_j & {\rm for} & i=j,\\
    \cos\theta\,\bsigma_j+\sin\theta\,\displaystyle\sum_{k=1}^3\epsilon_{ijk}\,\bsigma_k,
    &{\rm for} & i\neq j,\end{array}\right.
    \label{id2}
    \ee
to show that
    \bea
    \bs_1(\vartheta,\varphi)&=&(\sin^2\varphi+\cos\vartheta\cos^2\varphi)\bsigma_1
    -\sin\varphi\cos\varphi(1-\cos\vartheta)\bsigma_2-\sin\vartheta\cos\varphi\,\bsigma_3,
    \label{bs1=}
    \\
    \bs_2(\vartheta,\varphi)&=&-\sin\varphi\cos\varphi(1-\cos\vartheta)\bsigma_1+
    (\cos^2\varphi+\cos\vartheta\sin^2\varphi)\bsigma_2-\sin\vartheta\sin\varphi\,\bsigma_3,
    \label{bs2=}
    \\
    \bs_3(\vartheta,\varphi)&=&\sin\vartheta\cos\varphi\,\bsigma_1+\sin\vartheta\sin\varphi\,\bsigma_2+
    \cos\vartheta\,\bsigma_3=\hat x\cdot\vec\bsigma,
    \label{bs3=}
    \eea
where $\delta_{ij}$ and $\epsilon_{ijk}$ are respectively the kronecker delta and Levi Civita epsilon symbols,
    \bea
    &\hat x:=(x_1,x_2,x_2), \quad\quad\quad
    \vec\bsigma:=(\bsigma_1,\bsigma_2,\bsigma_3).&
    \label{hat-x=1}\\
    &x_1:=\sin\vartheta\cos\varphi,
    \quad\quad x_2:=\sin\vartheta\sin\varphi,
    \quad\quad  x_3:=\cos\vartheta,&
    \label{hat-x=2}
    \eea
and $\hat x\cdot\vec\bsigma:=\sum_{j=1}^3 x_j\bsigma_j$.

\begin{figure}
    \begin{center}
    \includegraphics[scale=.40]{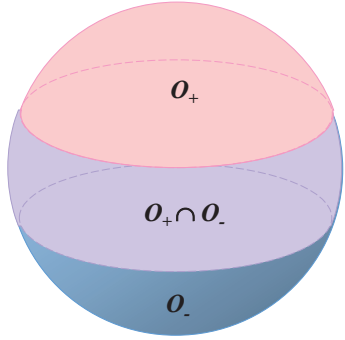}
    \caption{Covering of $S^2$ by a pair of local coordinate patches $\cO_\pm$.}
    \label{fig1}
    \end{center}
    \end{figure}

According to (\ref{bs=}) and (\ref{bs3=}),
    \be
    \bU[R]\bsigma_3\bU[R]^\dagger=\hat x\cdot\vec\bsigma.
    \label{U-id}
    \ee
Substituting (\ref{diag-eta=}) and (\ref{U-id}) in (\ref{matrix-eta=}) and simplifying the result give
    \be
    \boldsymbol{\eta}[R]=\chi_+\bI+\chi_-\,\hat x\cdot\vec\bsigma.
    \label{matrix-eta=2}
    \ee
This equation identifies the unit sphere, $S^2:=\big\{\hat x\in\R^3~\big|~|\hat x|=1\big\}$, as a natural choice for the base manifold of the vector bundle $\cE$. $\bcU(\vartheta,\varphi)$ is single-valued at every point of $S^2$ except the south pole $S$ (where $\vartheta=\pi$). This is actually related to the fact that we need at least two local coordinate patches to cover $S^2$. For definiteness we identify these with
    \begin{align*}
    &\cO_+:=\{ (\varphi,\vartheta)~|~\varphi\in[0,2\pi),~\vartheta\in[0,\vartheta_+)~\}
    \subsetneq S^2\setminus\{S\},\\
    &\cO_-:=\{ (\varphi,\vartheta)~|~\varphi\in[0,2\pi),~\vartheta\in(\vartheta_-,\pi]~\}        \subsetneq S^2\setminus\{N\},
    \end{align*}
where $\vartheta_\pm$ are a pair of angles such that $0<\vartheta_-<\vartheta_+<\pi$, and $N$ denotes the north pole of $S^2$. See Fig.~\ref{fig1}

Clearly we can take $R=(\vartheta,\varphi)$. In particular, the parameters $\xi$ and $\zeta$ are functions of $\vartheta$ and $\varphi$. They may be defined for $\vartheta\geq\vartheta_+$, but need not take a positive value for $\vartheta\geq\vartheta_+$. A typical example is: $\xi:=1$ and $\zeta:=1-\cos\vartheta/\cos\vartheta_+$.

For $R\in\cO_-$, we identify the metric operator $\tilde\eta[R]:\sH\to\sH$ with the one whose matrix representation in the basis $\{e_1,e_2\}$ takes the form
    \be
    \tilde{\boldsymbol{\eta}}[R]=\tilde\bU[R]\,\tilde{\boldsymbol{\eta}}_d[R]\tilde\bU[R]^\dagger
    =\tilde\chi_+\bI+\tilde\chi_-\,\hat{\tilde x}\cdot\vec\bsigma
    ,
    \label{matrix-tilde-eta=}
    \ee
where
    \begin{align}
    &\tilde\bU[R]:=\bcU(\pi-\vartheta,\varphi)=
    e^{-i\varphi\bsigma_3/2} e^{-i(\pi-\vartheta)\bsigma_2/2}e^{i\varphi\bsigma_3/2},
    \label{tilde-U=}\\[3pt]
    &\tilde{\boldsymbol{\eta}}_d[R]=\left[\begin{array}{cc}
    \tilde\xi^2 & 0 \\
    0 & \tilde\zeta^2\end{array}\right]=\tilde\chi_+\bI+\tilde\chi_-\bsigma_3,\\
    &\tilde\chi_\pm:=\frac{\tilde\xi^2\pm\tilde\zeta^2}{2},\quad\quad\quad
    \hat{\tilde x}:=(x_1,x_2,-x_3),
    \label{tilde-x}
    \end{align}
where $\tilde\xi$ and $\tilde\zeta$ are possibly $\vartheta$- and
$\varphi$-dependent positive real numbers, and we have made use of
(\ref{id2}).

For $R\in\cO_-\cap\cO_+$, we can  relate $\tilde{\boldsymbol{\eta}}[R]$ to ${\boldsymbol{\eta}}[R]$ using  (\ref{matrix-rep-eta-g-eta}), if we set
    \bea
    &&\bg_{\alpha\beta,R}=\bg_{+ -,R}:=\bU[R]\:\bg_d[R]\:\tilde\bU[R]^\dagger,
    \label{transition-matrix}\\[6pt]
    &&\bg_d[R]:=\left[\begin{array}{cc}
     \tilde\xi/\xi & 0 \\
    0 & \tilde\zeta/\zeta\end{array}\right]=\gamma_+\bI+\gamma_-\bsigma_3,
    \label{transition-matrix-d}\\[3pt]
    &&\gamma_\pm:=\frac{1}{2}\left(\frac{ \tilde\xi}{\xi}\pm
    \frac{ \tilde\zeta}{\zeta}\right).
    \label{gamma-pm=}
    \eea
To obtain a more explicit expression for $\bg_{+ -,R}$, we
substitute (\ref{U=}) and (\ref{tilde-U=}) in
(\ref{transition-matrix}) and make use of the identities (\ref{id2})
and
    \begin{align}
    &e^{i\theta\bsigma_i}=\cos\theta\,\bI+i\sin\theta\,\bsigma_j,
    &&\bsigma_i\bsigma_j=\delta_{ij}\bI+i\sum_{k=1}^3\epsilon_{ijk}\bsigma_k.
    \label{id1}
    \end{align}
This yields:
    \bea
    \bg_{+ -,R}&=&
    (\gamma_+\bI+\gamma_-\hat x\cdot\vec{\bsigma})\,\bsigma_3(\hat x'\cdot\vec{\bsigma})
    \label{g=x-prime}\\
    &=&
    \gamma_+\sin\vartheta\,\bI+
    (\gamma_-\cos\varphi-i\gamma_+\cos\vartheta\sin\varphi)\bsigma_1+
    (\gamma_-\sin\varphi+i\gamma_+\cos\vartheta\cos\varphi)\bsigma_2,
    \nn\\[6pt]
    &=&\left[\begin{array}{cc}
    \gamma_+\sin\vartheta&
    e^{-i\varphi}(\gamma_-+\gamma_+\cos\vartheta)\\[3pt]
    e^{i\varphi}(\gamma_--\gamma_+\cos\vartheta) &
    \gamma_+\sin\vartheta
    \end{array}\right],
    \label{bg=}
    \eea
where $\hat x'$ is the unit vector obtained from $\hat x$ by changing $\vartheta$ to $\frac{\pi}{2}-\vartheta$, i.e.,
    \[\hat x'=(\cos\vartheta\cos\varphi,\cos\vartheta\sin\varphi,\sin\vartheta).\]
Equations~(\ref{matrix-eta=2}), (\ref{matrix-tilde-eta=}),  and
(\ref{bg=}) give the matrix representation of the metric operators
$\eta[R]$ and $\tilde\eta[R]$, and the transition function
$g_{+-,R}$ in the basis $\{e_1,e_2\}$. We can use these to determine
the Hermitian vector bundle $\cE$ provided that we specify the
$\vartheta$- and $\varphi$-dependence of
$\xi,\zeta,\tilde\xi$, and $\tilde\zeta$.

\subsection{Construction of $\cA$}

In order to determine a metric-compatible connection $\cA$ on $\cE$, we need to give the expression for the local connection one-forms $\bA$ and $\tilde\bA$ that define parallel transportation in $\cO_+$ and $\cO_-$, respectively. According to (\ref{gen-A}), $\bA=\bA_0+\boldsymbol{\omega}$, where $\bA_0:=-i\bfeta^{-1}d\bfeta/2$ and $\boldsymbol{\omega}$ is an $\boldsymbol{\eta}$-pseudo-Hermitian one-form, i.e., its components satisfy (\ref{omega-ph}). Similarly, we have $\tilde\bA=\tilde\bA_0+\tilde\bomega$,  $\tilde\bA_0:=-i\tilde\bfeta^{-1}d\tilde\bfeta/2$, and $\tilde\bomega$ is a $\tilde\bfeta$-pseudo-Hermitian one-form.

To determine $\bA_0$, we substitute (\ref{matrix-eta=2}) in (\ref{A-zero}) and make use of (\ref{chi=}) and (\ref{id1}) to write the result in the form
    {\small \bea
    \bA_0[R]&=&-\frac{i}{2}\left\{\left(\frac{d\xi}{\xi}+\frac{d\zeta}{\zeta}\right)\bI+
    \left(\left[\frac{d\xi}{\xi}-\frac{d\zeta}{\zeta}\right]\hat x+
    \left[\frac{\xi^4-\zeta^4}{4\xi^2\zeta^2}\right]d\hat x
    -i\left[\frac{(\xi^2-\zeta^2)^2}{4\xi^2\zeta^2}\right]\hat x\times d\hat x\right)
    \cdot\bsigma\right\},~~
    \label{A-zero=}
    \eea}%
where $R\in\cO_+$ and $\times$ stands for the cross product. It is easy to show that
    \bea
    d\hat x&=&
    \big(\,\cos\vartheta\cos\varphi\,,\,
    \cos\vartheta\sin\varphi \,,\,
    -\sin\vartheta \,\big)\,d\vartheta+
    \big(-\sin\varphi\,,\,
    \cos\varphi\,,\,0\,\big)\sin\vartheta\, d\varphi,
    \label{dx=}\\
    \hat x\times d\hat x&=&
    \big(-\sin\varphi\,,\,
    \cos\varphi\,,\,0\big)\,d\vartheta-\mbox{\large$\frac{1}{2}$}
    \big(\sin(2\vartheta)\cos\varphi\,,\,
    \sin(2\vartheta)\sin\varphi\,,\,
    -1+\cos(2\vartheta)\big)\,d\varphi.
    \label{xdx=}
    \eea
These are manifestly single-valued for $0<\theta<\pi$. In Appendix~A, we show that their apparent multi-valuedness at $\theta=0$ is due to the multi-valuedness of the coordinates $(\vartheta,\varphi)$ at the north pole $N$, and that (\ref{dx=}) and (\ref{xdx=}) actually define $d\hat x$ and $\hat x\times d\hat x$ as well-defined functions in $S^2\setminus\{S\}$. Let us also note that according to (\ref{hat-x=1}), (\ref{dx=}), and (\ref{xdx=}),
    \bea
    \hat x\cdot\bsigma&=&\sin\vartheta\,\bfs_1(\varphi)+\cos\vartheta\,\bsigma_3,
    \label{x-sigma=}\\
    d\hat x\cdot\bsigma&=&[\cos\vartheta\,\bfs_1(\varphi)-\sin\vartheta\,\bsigma_3]
    d\vartheta+\bfs_2(\varphi)\sin\vartheta\,d\varphi,
    \label{dx-sigma=}\\
    (\hat x\times d\hat x)\cdot\bsigma&=&
    \bfs_2(\varphi) d\vartheta-[\cos\vartheta\,\bfs_1(\varphi)
    -\sin\vartheta\,\bsigma_3]\sin\vartheta\,d\varphi,
    \label{xdx-sigma=}
    \eea
where
    \begin{align}
    &\boldsymbol{\fs}_1(\varphi):=\cos\varphi\,\bsigma_1+\sin\varphi\,\bsigma_2,
    &&
    \boldsymbol{\fs}_2(\varphi):=-\sin\varphi\,\bsigma_1+\cos\varphi\,\bsigma_2.
    \label{fss=}
    \end{align}

The calculation of $\tilde\bA_0$ is similar. It is given by the right-hand side of (\ref{A-zero=}) with  $R\in\cO_-$ and $\xi$, $\zeta$, and $\hat x$ respectively replaced with $\tilde\xi$, $\tilde\zeta$, and $\hat{\tilde x}$, i.e.,
    {\small \bea
    \tilde\bA_0[R]&=&
    -\frac{i}{2}\left\{\left(\frac{d\tilde\xi}{\tilde\xi}+
    \frac{d\tilde\zeta}{\tilde\zeta}\right)\bI+
    \left(\left[\frac{d\tilde\xi}{\tilde\xi}-\frac{d\tilde\zeta}{\tilde\zeta}\right]\hat{\tilde x}+
    \left[\frac{\tilde\xi^4-\tilde\zeta^4}{4\tilde\xi^2\tilde\zeta^2}\right]d\hat{\tilde x}
    -i\left[\frac{(\tilde\xi^2-\tilde\zeta^2)^2}{4\tilde\xi^2\tilde\zeta^2}\right]
    \hat{\tilde x}\times d\hat{\tilde x}\right)
    \cdot\bsigma\right\}.~~
    \label{tA-zero=}
    \eea}

Next, we recall that in $\cO_-\cap\cO_+$, $\tilde\bA$, $\bA$, and $\bg_{+ -,R}$ fulfil (\ref{connection-trans}). Expressing $\bA$ and $\tilde\bA$ in terms of $\bomega$ and $\tilde\bomega$ in this equation and making use of $\bA_0:=-i\bfeta^{-1}d\bfeta/2$ and  $\tilde\bA_0:=-i\tilde\bfeta^{-1}d\tilde\bfeta/2$, we find
    \be
    \tilde{\boldsymbol{\omega}}=\bg^{-1}\left\{\boldsymbol{\omega}
    -\frac{i}{2}\left[d\bg\,\bg^{-1}-\boldsymbol{\eta}^{-1}(d\bg\,\bg^{-1})^\dagger
    \boldsymbol{\eta}\right]\right\}\bg,
    \label{tomega-omega}
    \ee
where $\bg$ abbreviates $\bg_{+-,R}$.

To explore the consequences of (\ref{tomega-omega}), we first note that every $\boldsymbol{\eta}$-pseudo-Hermitian matrix $\bM$ admits a decomposition of the form $\boldsymbol{\rho}^{-1}\bM_H\,\boldsymbol{\rho}$, where $\boldsymbol{\rho}:=\sqrt{\boldsymbol{\eta}}$ is the positive square root of $\boldsymbol{\eta}$, and $\bM_H$ is a Hermitian matrix \cite{jpa-2003}.  This implies the existence of Hermitian matrix-valued one-forms $\bomega_H$ and $\tilde\bomega_H$ satisfying
    \begin{align}
    &\bomega=\boldsymbol{\rho}^{-1}\bomega_H\,\boldsymbol{\rho},
    &&\tilde\bomega=\tilde\brho^{-1}\tilde\bomega_H\,\tilde\brho,
    \label{omegas=}
    \end{align}
where $\tilde\brho:=\sqrt{\tilde\bfeta}$. Multiplying both sides of  (\ref{tomega-omega}) by $\tilde{\brho}$ from the left and $\tilde{\brho}^{-1}$ from the right, we find
    \be
    \tilde{\boldsymbol{\omega}}_H=\bcG^{-1}\left(\boldsymbol{\omega}_H+\bGamma\right)\bcG,
    \label{tomega=omega}
    \ee
where $\boldsymbol{\cG}$ denotes the matrix representation of the transition function $\cG_{+-,R}$ of the bundle $u(\cE)$ in the basis $\{e_1,e_2\}$, i.e.,
    \be
    \boldsymbol{\cG}:=\boldsymbol{\rho} \:\bg \:
    \tilde{\boldsymbol{\rho}}^{-1},
    \label{bcG-def}
    \ee
and
    \be
    \bGamma:=
    -\frac{i}{2}\left[\brho\,d\bg\:\bg^{-1}\brho^{-1}-\left(\brho\,d\bg\:\bg^{-1}\brho^{-1}\right)^\dagger\right].
    \label{Gamma-def}
    \ee
Notice that $\boldsymbol{\cG}$ and $\bGamma$ are respectively a unitary matrix-valued function and a Hermitian matrix-valued one-form defined in $\cO_+\cap\cO_-$.

To determine the $\vartheta$- and $\varphi$-dependence of $\bcG$, we need to compute $\brho$ and $\tilde\brho$. We do this using the method we employed for calculating $\boldsymbol{\eta}$. The outcome is:
    \bea
    \boldsymbol{\rho}[R]&=&\bU[R]\:\brho_d[R]\:\bU[R]^\dagger=
    \varrho_+\bI+\varrho_-\,\hat x\cdot\vec\bsigma,\quad\quad R\in\cO_+,
    \label{matrix-rho=2}\\
    \tilde{\brho}[R]&=&\tilde\bU[R]\:\tilde\brho_d[R]\:\tilde\bU[R]^\dagger=
    \tilde\varrho_+\bI+\tilde\varrho_-\,\hat{\tilde x}\cdot\vec\bsigma,\quad\quad R\in\cO_-,
    \label{trho=U-rho-U}
    \eea
where
    \begin{align}
    &\brho_d[R]:=\left[\begin{array}{cc}
    \xi & 0 \\
    0 & \zeta\end{array}\right],
    &&\varrho_\pm:=\frac{\xi\pm\zeta}{2},
    &&\tilde\brho_d[R]:=\left[\begin{array}{cc}
    \tilde\xi & 0 \\
    0 & \tilde\zeta\end{array}\right],
    &&\tilde\varrho_\pm:=\frac{\tilde\xi\pm\tilde\zeta}{2}.
     \label{eps=}
    \end{align}
Substituting (\ref{transition-matrix}), (\ref{matrix-rho=2}), and (\ref{trho=U-rho-U}) in (\ref{bcG-def}), we have
    \be
    \bcG 
    =\bU[R] \:\tilde\bU[R]^\dagger.
    \label{bcG=}
    \ee
Comparing this equation with (\ref{transition-matrix}), we see that $\bcG$ is given by the right-hand side of (\ref{bg=}) provided that we replace $\gamma_+$ and $\gamma_-$ with $1$ and $0$, respectively. In particular,
    \be
    \bcG=\bsigma_3(\hat x'\cdot\vec\bsigma)=
    \left[\begin{array}{cc}
        \sin\vartheta&e^{-i\varphi}\cos\vartheta\\[3pt]
        -e^{i\varphi} \cos\vartheta &\sin\vartheta\end{array}\right].
        \label{bcG=3}
    \ee
Observe that unlike $\bg$, $\bcG$ does not involve $\xi$, $\zeta$, $\tilde\xi$, and $\tilde\zeta$, and that it is single-valued in $S^2\setminus\{N,S\}$.

Next, we use (\ref{gamma-pm=}), (\ref{bg=}), and (\ref{Gamma-def}) -- (\ref{eps=}) to
compute $\bGamma$. This yields
    \bea
    \bGamma
    &=&\cX\,\bGamma_+(\vartheta,\varphi)+\tilde\cX\,\bGamma_-(\vartheta,\varphi)
    +\bGamma_0(\vartheta,\varphi),
    \label{Gamma=123}
    \eea
where
    \begin{align}
    &\cX:=\frac{\xi^2+\zeta^2}{4\xi\zeta},
    \quad\quad\quad
    \tilde\cX:=\frac{\tilde\xi^2+\tilde\zeta^2}{4\tilde\xi\tilde\zeta},
    \label{X=tX}\\
    &\bGamma_+(\vartheta,\varphi):=-\bfs_2(\varphi)\,d\vartheta+
    [\cos\vartheta\,\bfs_1(\varphi)-\sin\vartheta\,\bsigma_3]
    \sin\vartheta\,d\varphi,
    \label{bGamma+}\\
    &\bGamma_-(\vartheta,\varphi):=
    -\bfs_2(\varphi)\,d\vartheta-
    [\cos\vartheta\,\bfs_1(\varphi)-\sin\vartheta\,\bsigma_3]
    \sin\vartheta\,d\varphi,
    \label{bGamma-}\\
    &\bGamma_0(\vartheta,\varphi):=
    [-\sin\vartheta\,\bfs_1(\varphi)+\cos\vartheta\,\bsigma_3]\cos\vartheta\,d\varphi,
    \label{bGamma0}
    \end{align}
and $\bfs_\ell$ are given by (\ref{fss=}). As we explain in Appendix~A, $\sin\varphi\,d\vartheta$ and $\cos\varphi\,d\vartheta$ are single-valued at the poles of $S^2$ and hence in $S^2$. Therefore, the same holds for $\bfs_2\,d\vartheta$ and consequently $\bGamma_\pm$ and $\bGamma_0$.

Eqs.~(\ref{tomega=omega}) and (\ref{Gamma=123}) together with the fact that $\bomega_H$ and $\tilde\bomega_H$ are respectively defined in $\cO_+$ and $\cO_-$ suggest expressing $\bomega_H$ in the form
    \be
    \bomega_H=\balpha(\vartheta,\varphi)-\cX\,\bGamma_+(\vartheta,\varphi)
    -{\bGamma}_0(\vartheta,\varphi),
    \label{omega-H=131}
    \ee
where $\balpha(\vartheta,\varphi)$ is a Hermitian matrix-valued one-form defined in $S^2$. By virtue of (\ref{tomega=omega}) and (\ref{Gamma=123}), this equation allows us to write
    \bea
    \tilde\bomega_H&=&\tilde\balpha(\vartheta,\varphi)+
    \tilde\cX\,\tilde{\bGamma}_-(\vartheta,\varphi),
    \label{tomega-H=131}
    \eea
where
    \begin{align}
    &\tilde\balpha:=\bcG^{-1}\balpha\,\bcG,
    &&\tilde{\bGamma}_-:=\bcG^{-1}{\bGamma}_-\,\bcG.
    \label{new-defs-GS}
    \end{align}
In Appendix~B we establish the following remarkable identity
    \be
    \tilde\bGamma_-(\vartheta,\varphi)=-\bGamma_+(\pi-\vartheta,\varphi).
    \label{r-identity}
    \ee
Using this in (\ref{tomega-H=131}) we find
    \be
    \tilde\bomega_H=\tilde\alpha(\vartheta,\varphi)-
    \tilde\cX\,{\bGamma}_+(\pi-\vartheta,\varphi).
    \label{tomega-H=132}
    \ee

Because $\bGamma_0$, $\bGamma_+$, and $\balpha$ are single-valued in $S^2$, (\ref{omega-H=131}) and (\ref{tomega-H=132}) identify $\bomega_H$ and $\tilde\bomega_H$ as well-defined Hermitian matrix-valued one-forms in $\cO_+$ and $\cO_-$, respectively. Substituting these in (\ref{omegas=}), we obtain valid expressions for $\bomega$ and $\tilde\bomega$ which together with $\bA_0$ and $\tilde\bA_0$ of Eqs.~(\ref{A-zero=}) and (\ref{tA-zero=}) yield the local connection one-forms $\bA$ and $\tilde\bA$ via
    \begin{align}
    &\bA=\bA_0+\bomega,
    &&\tilde\bA=\tilde\bA_0+\tilde\bomega.
    \label{AA=}
    \end{align}

\subsection{Dynamics of the system}
\label{Sec4-3}

In order to formulate the dynamics of the system we use its local Hermitian representations $(\sH,h)$ and $(\sH,\tilde h)$ which are respectively valid in the local coordinate patches $\cO_+$ and $\cO_-$. Following the standard practice, we identify linear operators acting in $\sH$ with their matrix representation in the basis $\{e_1,e_2\}$. Then in view of (\ref{H-ph=}) and (\ref{h=2}), we can write the matrix representation of $h(t)$ and $\tilde h(t)$ in the form
    \bea
    \bh(t)&=&\bh_E(t)+\bh_\omega(t)+\bh_\rho(t),
    \label{h=21}\\
    \tilde \bh(t)&=&\tilde \bh_E(t)+\tilde\bh_{\tilde\omega}(t)+\tilde\bh_{\tilde\rho(t)},
    \label{th=21}
    \eea
where $\bh_E(t)=\bh_E[R(t)]$ and $\tilde\bh_E(t)=\tilde\bh_E[R(t)]$ respectively represent the energy observables in $\cO_+$ and $\cO_-$,
    \begin{align}
    &\bh_\omega(t):=\dot R^a(t)\bomega_{Ha}[R(t)],
    &&\tilde\bh_{\tilde \bomega}(t):=\dot R^a(t)\tilde\bomega_{Ha}[R(t)],
    \label{bhs=}
    \end{align}
$\bomega_{Ha}$ and $\tilde\bomega_{Ha}$ respectively stand for the components of the Hermitian matrix-valued one-forms $\bomega_{H}$ and $\tilde\bomega_{H}$,
    \bea
    \bh_\rho(t)&:=&\frac{i}{2}\left[\dot\brho(t),\brho(t)^{-1}\right]=\dot R^a(t)\btheta_a[R(t)],
    \label{bh-rho=}\\
    \tilde\bh_{\tilde\rho}(t)&:=&\frac{i}{2}\left[\dot{\tilde\brho}(t),\tilde{\brho}(t)^{-1}\right]=\dot R^a(t)\tilde\btheta_a[R(t)],
    \label{tbh-rho=}
    \eea
and $\btheta_a$ and $\tilde\btheta_a$ are the components of the Hermitian matrix-valued one-forms:
    \begin{align}
    &\btheta:=\frac{i}{2}\left[d\brho,\brho^{-1}\right],
    &&\tilde\btheta:=\frac{i}{2}\left[d\tilde\brho,\tilde\brho^{-1}\right].
    \label{tau-ttau-def}
    \end{align}

A choice for $\bh_E[R]$ and $\tilde\bh_E[R]$ that  fulfills
    \be
    \tilde\bh_E[R]=\bcG^{-1}\bh_E[R]\,\bcG
    \label{th=4}
    \ee
for all $R\in\cO_-\cap\cO_+$ is equivalent to a choice of a global section $\fh_E$ of the real vector bundle $\fu(\cE)$ that specifies the energy observable for the system.

Without loss of generality, we can take
     \be
    \bh_E[R]:=\frac{\varepsilon}{2}\:\hat y\cdot\vec\bsigma,
    \label{energy=}
    \ee
where $\varepsilon$ is a possibly $R$-dependent real parameter, and $\hat y=(y_1,y_2,y_3)\in\R^3$ is a unit vector whose components $y_j$ are smooth functions of $R\in\cO_+$. In other words, $\varepsilon$ and $\hat y$ are respectively smooth functions mapping $\cO_+$ to $\R$ and $S^2$. In view of  (\ref{th=4}) and (\ref{energy=}),
    \bea
    \tilde\bh_E[R]&=&\frac{\varepsilon}{2}\:\hat y\cdot\vec{\tilde\bsigma},
    \label{th-E=}
    \eea
where $\vec{\tilde\bsigma}=(\tilde\bsigma_1, \tilde\bsigma_2, \tilde\bsigma_3)$ and
$\tilde\bsigma_j:=\bcG^{-1}\bsigma_j\,\bcG$. Substituting (\ref{bcG=3}) in the latter relation, we have
    \bea
    \tilde\bsigma_1&=&(\sin^2\vartheta-\cos^2\vartheta\cos 2\varphi)\bsigma_1
    -\cos^2\vartheta \sin 2\varphi\,\bsigma_2
    -\sin 2\vartheta \cos\varphi\,\bsigma_3,
    \label{gsg1=}\\
    \tilde\bsigma_2&=&-\cos^2\vartheta \sin 2\varphi\,\bsigma_1
    +(\sin^2\vartheta+\cos^2\vartheta\cos 2\varphi)\bsigma_2
    -\sin 2\vartheta \sin \varphi\,\bsigma_3,
    \label{gsg2=}\\
   \tilde\bsigma_3&=&\sin 2\vartheta(\cos\varphi\,\bsigma_1+\sin\varphi\,\bsigma_2)
   -\cos 2\vartheta\,\bsigma_3=\sin 2\vartheta\,\bfs_1(\varphi)-\cos 2\vartheta\,\bsigma_3.
    \label{gsg3=}
    \eea
Equation~(\ref{th-E=}) gives $\tilde\bh_E[R]$ in $\cO_+\cap\cO_-$. To determine $\tilde\bh_E[R]$ in $\cO_-\setminus\cO_+$ we need to extend the definition of $\varepsilon$ and $\hat y$ to $S^2$ in a smooth manner. We also note that the right-hand sides of (\ref{gsg1=}) and (\ref{gsg2=}) are not single-valued at the poles of $S^2$ where $\vartheta\in\{0,\pi\}$. In particular, Eqs.~(\ref{th-E=}) -- (\ref{gsg3=}) determine $\tilde\bh_E[R]$ in $\cO_-$ provided that we choose a particular value for $\tilde\bsigma_1$ and $\tilde\bsigma_2$ at the south pole.

Next, we examine $\bh_\omega(t)$ and $\tilde\bh_{\tilde \bomega}(t)$. According to (\ref{bhs=}), they are determined by the $\bomega_{H}$ and $\tilde\bomega_{H}$ of Eqs.~(\ref{omega-H=131}) and (\ref{tomega-H=132}). These involve the Hermitian matrix-valued one-forms $\balpha$ and $\tilde\balpha$. It is not difficult to perform a gauge transformation that discards the trace of $\balpha$. This means that, without loss of generality, we can express $\balpha$ in the form
    \be
    \balpha=\vec\alpha\cdot\vec\bsigma=
    \left(\vec\alpha_{\vartheta}\, d\vartheta+
        \vec\alpha_{\varphi}\, d\varphi\right)\cdot\vec\bsigma,
    \label{alpha=1}
    \ee
where $\vec\alpha:=(\alpha_1,\alpha_2,\alpha_3)$, $\alpha_j$ are one-forms defined in $S^2$ and having  components $\alpha_{j \vartheta}$ and $\alpha_{j \varphi}$, $\vec\alpha_{\vartheta}:=(\alpha_{1\vartheta},
\alpha_{2\vartheta},\alpha_{3\vartheta})$, and $\vec\alpha_{\varphi}:=(\alpha_{1\varphi},
\alpha_{2\varphi},\alpha_{3\varphi})$. In view of (\ref{new-defs-GS}) and (\ref{alpha=1}),
    \be
    \tilde\balpha=\vec\alpha\cdot\vec{\tilde\bsigma}=
    \left(\vec\alpha_{\vartheta}\, d\vartheta+
        \vec\alpha_{\varphi}\, d\varphi\right)\cdot\vec{\tilde\bsigma}.
    \label{talpha=1}
    \ee
Substituting (\ref{omega-H=131}) and (\ref{tomega-H=132}) in (\ref{bhs=}) and making use of (\ref{X=tX}),(\ref{bGamma+}), (\ref{alpha=1}), and (\ref{talpha=1}), we have
    \bea
    \bh_\omega(t)&=&\frac{\xi^2+\zeta^2}{4\xi\zeta}
    \left\{\bfs_2(\varphi)\dot\vartheta-
    \big[\cos\vartheta\,\bfs_1(\varphi)-\sin\vartheta\,\bsigma_3\big]
    \sin\vartheta\,\dot\varphi\right\}+\nn\\
    &&\big[\sin\vartheta\,\bfs_1(\varphi)
    -\cos\vartheta\,\bsigma_3\big]\cos\vartheta\,\dot\varphi+
    \left(\vec\alpha_{\vartheta}\, \dot\vartheta+
    \vec\alpha_{\varphi}\, \dot\varphi\right)\cdot\vec{\bsigma},
    \label{h-omega=}\\
    \tilde\bh_{\tilde\omega}(t)&=&\frac{\tilde\xi^2+\tilde\zeta^2}{4\tilde\xi\tilde\zeta}
    \left\{-\bfs_2(\varphi)\dot\vartheta+
    \big[\cos\vartheta\,\bfs_1(\varphi)+\sin\vartheta\,\bsigma_3\big]
    \sin\vartheta\,\dot\varphi\right\}+
    \left(\vec\alpha_{\vartheta}\, \dot\vartheta+
    \vec\alpha_{\varphi}\, \dot\varphi\right)\cdot\vec{\tilde\bsigma}.
    \label{th-omega=}
    \eea

In Appendix~C, we outline a derivation of $\bh_\rho$ and $\tilde\bh_{\tilde\rho}$ which yields
    \bea
    \bh_\rho(t)&=&\frac{(\xi-\zeta)^2}{4\xi\zeta}\Big\{-\bfs_2(\varphi)\,\dot\vartheta+
    \big[\cos\vartheta\,\bfs_1(\varphi)-\sin\vartheta\,\bsigma_3\big]\sin\vartheta\,\dot\varphi\Big\}.
    \label{bh-rho=11}\\
    \tilde\bh_{\tilde\rho}(t)&=&
    \frac{(\tilde\xi-\tilde\zeta)^2}{4\tilde\xi\tilde\zeta}
    \Big\{\bfs_2(\varphi)\,\dot\vartheta-
    \big[\cos\vartheta\,\bfs_1(\varphi)+\sin\vartheta\,\bsigma_3\big]\sin\vartheta\,\dot\varphi\Big\}.
    \label{tbh-rho=11}
    \eea

According to Eqs.~(\ref{h=21}), (\ref{th=21}), (\ref{energy=}),
(\ref{th-E=}), (\ref{h-omega=}), (\ref{th-omega=}),
(\ref{bh-rho=11}), and (\ref{tbh-rho=11}),  the dynamics of our
quantum system is determined by the following Hermitian matrix Hamiltonians
in $\cO_+$ and $\cO_-$, respectively.
    \bea
    \bh(t)&=&\left(\frac{\varepsilon}{2}\:\hat y+ \vec\alpha_{\vartheta}\, \dot\vartheta+
        \vec\alpha_{\varphi}\, \dot\varphi\right)\cdot\vec{\bsigma}+\frac{1}{2}\Big\{
    \bfs_2(\varphi)\,\dot\vartheta+\big[-\cos\vartheta\,\bfs_1(\varphi)+\sin\vartheta\,\bsigma_3\big]\sin\vartheta\,
    \dot\varphi\Big\}+\nn\\
    &&
    \big[\sin\vartheta\,\bfs_1(\varphi)-\cos\vartheta\,\bsigma_3\big]\cos\vartheta\,
    \dot\varphi,
    \label{bh=final}\\[6pt]
    \tilde\bh(t)&=&\left(\frac{\varepsilon}{2}\:\hat y+ \vec\alpha_{\vartheta}\, \dot\vartheta+
        \vec\alpha_{\varphi}\, \dot\varphi\right)\cdot\vec{\tilde\bsigma}+
   \frac{1}{2}\Big\{-\bfs_2(\varphi)\,\dot\vartheta+
   \big[\cos\vartheta\,\bfs_1(\varphi)+\sin\vartheta\,\bsigma_3\big]\sin\vartheta\,
    \dot\varphi\Big\}.
    \label{tbh=final}
    \eea
As seen from these relations the variables $\xi,\zeta,\tilde\xi$, and $\tilde\zeta$ drop out of the expression for $\bh(t)$ and $\tilde\bh(t)$. In particular, the right-hand sides of (\ref{bh=final}) and (\ref{tbh=final}) are well-defined in $S^2$. This is clearly not true for the Hamiltonians $\bH(t)$ and $\tilde\bH(t)$ whose expression involves
$(\xi,\zeta)$ and $(\tilde\xi,\tilde\zeta)$, respectively.


\section{Summary and concluding remarks}

A consistent formulation of the dynamics of a quantum system with a
time-dependent state space requires the mathematical tools of the
theory of Hermitian vector bundles. In this study we employ these
tools to provide a satisfactory way of dealing with the no-go
theorem of Ref.~\cite{plb-2007}, elucidate the meaning of the energy
observable for such a system, and develop a geometric extension of
quantum mechanics that identifies the standard quantum mechanics
with local representations of a quantum theory defined in terms of a
Hermitian vector bundle $\cE$ endowed with a metric-compatible
connection $\cA$. In this theory a quantum system is determined by a
curve $\cC$ in the base manifold of $\cE$ and an energy observable
$\fh$ that together with $\cA$ specify the dynamics of the system.
The observables are global sections of a real vector bundle
$\fu(\cE)$ which is uniquely determined by $\cE$.

If we can find a local  trivialization $(\cO_\alpha, f_\alpha$) of
$\cE$ such that $\cC$ lies in $\cO_\alpha$, then the whole
description of the system can be achieved using the standard
formulation of quantum mechanics. In particular, we can define a
Hermitian Hamiltonian operator $h:\sH\to\sH$ that carries the
information of the restrictions of $\cA$ and $\fh$ to $\cC$ and acts
in a standard time-independent Hilbert space $\sH$. The Hamiltonian
$h$ is the sum of two parts:
    \begin{enumerate}
    \item a geometric part $h_{A}$ that reflects the properties the connection $\cA$ and is responsible for the horizontal evolutions of the system;
    \item a nongeometric part $h_E$ that gives the energy observable for the system and generates the vertical evolutions of the system.
    \end{enumerate}

If $h_E=0$, the evolution  determined by the Hamiltonian $h$ via the
standard Schr\"odinger equation is purely geometrical. This means
that the evolving state depends solely on the curve $\cC$ and the
geometry of the bundle $\cE$. In particular, it is
time-reparametrization-invariant. In general $h_E\neq 0$, and the
evolution of the system involves geometric as well as nongeometric
ingredients. For example, if we examine adiabatic changes of the
system and perform Berry's investigation \cite{berry-1984} of the
evolution of an eigenstate of the initial value of the Hamiltonian
$h$, we find that due to the presence of the geometric part of the
Hamiltonian $h$, the standard dynamical phase acquired by the state
vector contains a geometric part! This is easy to see for the
two-level systems we have examined in Sec.~\ref{S4}.

The geometric extension of  quantum mechanics that we propose in
this article introduces a topological character to it without
affecting its measurement theoretical basis. This has to do with the
fact that the underlying Hermitian vector bundle $\cE$ can have a
nontrivial topology. The information about the topology of $\cE$
does not seem to affect the evolution of a single quantum system,
for in general the restriction of $\cE$ to the curve $\cC$ yields a
trivial vector bundle on $\cC$. The situation is analogous to the
geometric setups where adiabatic geometric phases are identified
with the holonomies of certain line bundles \cite{book2,simon}. The
topology of these line bundles do not have a direct effect on the
geometric phase acquired by a single evolving state \cite{jpa-1993}.
It is, however, the opinion of the author that introducing
topological features to quantum theory in a fundamental level may
lead to valuable developments towards the solution of some of the
most basic problems of theoretical physics. This is in line with the
teachings of Bryce and Cecile DeWitt to whose memory this work is
dedicated.

\vspace{12pt}
\noindent {\bf Note:} After the completion of this project, I was informed
of Ref.~\cite{moore} where the author considers a family of quantum systems
parameterized by points of the base manifold $M$ of a Hilbert bundle. The
fiber over a point $R\in M$ serves as the Hilbert space for the system
associated with $R$, and the observables are determined by global sections
of the bundle of bounded Hermitian operators acting in the Hilbert bundle.
For the case that the state space of these systems is finite-dimensional,
the Hilbert bundle and the bundle providing the observables coincide with
the vector bundles $\cE$ and $\fu(\cE)$ that we employ in this article.
There is however a major difference between the two approaches. The author
of Ref.~\cite{moore} does not identify the evolution of a quantum system
having a time-dependent Hilbert space with a lift of a curve 
in $M$ to $\cE$. Therefore in his formulation there is no need for the
introduction of a metric compatible connection on $\cE$ and a discussion
of the role and meaning of the energy observable for such a system.

\subsection*{Acknowledgments}
I am grateful to Farhang Loran for examining the first draft of this
manuscript and making constructive comments, and to Dieter Van den
Bleeken for bringing Ref.~\cite{moore} to my attention. This work has
been supported by the Turkish Academy of Sciences (T\"UBA).

\section*{Appendix~A: Single-valuedness of $d\hat x$ and $d\hat{\tilde x}$}

In this appendix we examine the behaviour of $d\hat x$ and
$d\hat{\tilde x}$  at the poles of $S^2$. Setting $\vartheta=0$ in
(\ref{dx=}) gives
    \be
    d\hat x\big|_{\vartheta=0}=(\cos\varphi\,d\vartheta,\sin\varphi\,
    d\vartheta,0)\big|_{\vartheta=0}.
    \label{dx=N}
    \ee
Because the value of $\varphi$ for $\vartheta=0$, this seems to
indicate  that $d\hat x$ and consequently $\hat x\times d\hat x$ are
multi-valued at the north pole $N$ of $S^2$.  In this appendix we
show that this is due to the multi-valuedness of the coordinates
$(\vartheta,\varphi)$ at $N$. Therefore it can be avoided if we use
a coordinate system in which every point of $S^2$ except the south
pole $S$ is determined by a unique pair of coordinates.  For example
consider following pair of Cartesian coordinates:
    \begin{align}
    &x:=\vartheta\cos\varphi,
    &&y:=\vartheta\sin\varphi,
    \label{cartesian}
    \end{align}
with $\vartheta\in[0,\pi)$ and $\varphi\in[0,2\pi)$. Clearly, each
point  of $S^2\setminus\{S\}$ is specified by a unique pair $(x,y)$.
In particular, $(x,y)$ corresponds to $N$ if and only if $x=y=0$. We
wish to compute $d\hat x$ and $\hat x\times d\hat x$ in the
$(x,y)$-coordinates.

According to (\ref{cartesian}),
$dx=\cos\varphi\,d\vartheta-\vartheta\sin\varphi\,d\varphi$ and
$dy=\sin\varphi\,d\vartheta+\vartheta\cos\varphi\,d\varphi$. In
particular,
$dx\big|_{\vartheta=0}=\cos\varphi\,d\vartheta\big|_{\vartheta=0}$
and
$dy\big|_{\vartheta=0}=\sin\varphi\,d\vartheta\big|_{\vartheta=0}$.
Comparing these equations with (\ref{dx=N}) we see that
    \be
    d\hat x\big|_{\vartheta=0}=(dx,dy,0)\big|_{\vartheta=0}.
    \label{dx=N2}
    \ee
This in turn implies that $d\hat x$ is indeed a well-defined
vector-valued one-form in $S^2\setminus\{S\}$. The same holds for
$\hat x\times d\hat x$, because $\hat x$ is a well-defined
vector-valued function in $S^2\setminus\{S\}$. In light of
(\ref{hat-x=1}), (\ref{hat-x=2}), and (\ref{dx=N2}), we have
    \be
    (\hat x\times d\hat x)\big|_{\vartheta=0}=(-dy,dx,0)\big|_{\vartheta=0}.
    \label{xdx=N2}
    \ee

Using a similar analysis we can show that $d\hat{\tilde x}$ and
$\hat{\tilde x}\times d\hat{\tilde x}$ are singule-valued functions
in $\cO_-$. To do this we introduce $\tilde\vartheta:=\pi-\vartheta$
and note that  $\hat{\tilde
x}:=(x_1,x_2,-x_3)=(\sin\tilde\vartheta\cos\varphi,\sin\tilde\vartheta\sin\varphi,\cos\tilde\vartheta)$.
The single-valuedness of $d\hat{\tilde x}$ and $\hat{\tilde x}\times
d\hat{\tilde x}$ in $\cO_-$ follows from the above analysis if we
use $\tilde\vartheta$ to play the role of $\vartheta$. In
particular, $d\hat{\tilde x}$ and $\hat{\tilde x}\times d\hat{\tilde
x}$ take the following value at the south pole $S$:
    \begin{align*}
    &d\hat{\tilde x}\Big|_{\vartheta=\pi}=(d\tilde x,d\tilde y,0)\Big|_{\tilde\vartheta=0},
    && (\hat{\tilde x}\times d\hat{\tilde x})\Big|_{\vartheta=\pi}=(-d\tilde y,d\tilde x,0)\Big|_{\tilde\vartheta=0},
    \end{align*}
where $\tilde x:=\tilde\vartheta\cos\varphi$, $\tilde y:=\tilde\vartheta\sin\varphi$, $\tilde\vartheta\in[0,\pi)$, and $\varphi\in[0,2\pi)$.

\section*{Appendix~B: Calculation of $\tilde\bGamma_-$}

In this appendix we derive an explicit expression for
$\tilde\bGamma_-$ that appears in the calculation of $\bh_\omega$ in
Sec.~\ref{Sec4-3}. First, we introduce
    \begin{align}
    &\tilde\bfs_\ell:=\bcG^{-1}\bfs_\ell\,\bcG,
    \label{new-defs}
    \end{align}
where $\bfs_\ell$ is given by (\ref{fss=}) and $\ell\in\{1,2\}$.  To
compute $\tilde\bfs_\ell$ we recall that
$\tilde\bsigma_j:=\bcG^{-1}\bsigma_j\,\bcG$ and use (\ref{fss=}),
(\ref{gsg1=}) -- (\ref{gsg3=}), and (\ref{new-defs}) to establish:
    \begin{align}
    &\tilde\bfs_1=\cos\varphi\,\tilde\bsigma_1+\sin\varphi\,\tilde\bsigma_2=
    -[\cos2\vartheta\,\bfs_1(\varphi)+\sin2\vartheta\,\bsigma_3],
    \label{tdf1=1}\\
    &\tilde\bfs_2=-\sin\varphi\,\tilde\bsigma_1+\cos\varphi\,\tilde\bsigma_2=\bfs_2(\varphi).
    \label{tdf2=1}
    \end{align}
These in turn imply
    $\cos\vartheta\,\tilde\bfs_1-\sin\vartheta\,\tilde\bsigma_3=
    -(\cos\vartheta\,\bfs_1+\sin\vartheta\,\bsigma_3)$.
In view of this relation and Eqs.~(\ref{bGamma-}), 
(\ref{new-defs-GS}), (\ref{new-defs}), and (\ref{tdf2=1}),
    \begin{align}
    \tilde\bGamma_-(\vartheta,\varphi)& =
    -\tilde\bfs_2(\vartheta,\varphi) \,d\vartheta-
    [\cos\vartheta\,\tilde\bfs_1(\vartheta,\varphi) -\sin\vartheta\,\tilde\bsigma_3(\vartheta,\varphi)]
    \sin\vartheta\,d\varphi\nn\\
    &=-\bfs_2(\varphi)\,d\vartheta+
    [\cos\vartheta\,\bfs_1(\varphi)+\sin\vartheta\,\bsigma_3]
    \sin\vartheta\,d\varphi.
    \label{tbgamma-=2}
    \end{align}
Comparing (\ref{tbgamma-=2}) with (\ref{bGamma+}) we obtain
(\ref{r-identity}).

\section*{Appendix~C: Calculation of $\bh_\rho$ and $\tilde\bh_{\tilde\rho}$}

In this appendix we summarize the calculation of the $\bh_\rho$ and
$\tilde\bh_{\tilde\rho}$ that enter the expression for the Hermitian
Hamiltonians $\bh$ and $\tilde\bh$ of Sec.~\ref{Sec4-3}. First, we
use (\ref{matrix-rho=2}), (\ref{bcU=}) -- (\ref{U-dagger=}), and
(\ref{id2}) to establish
    \bea
    [d\brho,\brho^{-1}]&=&\bU\left(2\bU^\dagger d\bU-\brho_d\bU^\dagger
    d\bU\brho_d^{-1}-
    \brho_d^{-1} \bU^\dagger d\bU\brho_d\right)\bU^\dagger,
    \label{bracket1}\\
    \bU^\dagger d\bU&=&\sum_{j=1}^3\beta_j\,\bsigma_j,
    \label{UdU=}
    \eea
where $\bU$ abbreviates $\bU[R]$, and
    \bea
    \beta_1&:=&\frac{i}{2}(\sin\vartheta\cos\varphi\,d\varphi+
    \sin\varphi\,d\vartheta),
    \label{beta1}\\
    \beta_2&:=&\frac{i}{2}(\sin\vartheta\sin\varphi\,d\varphi-\cos\varphi\,d\vartheta),
    \label{beta2}\\
    \beta_3&:=&\frac{i}{2}(1-\cos\vartheta)d\varphi.
    \label{beta3}
    \eea
Substituting (\ref{UdU=}) in (\ref{bracket1}) gives
    \be
    [d\brho,\brho^{-1}]=\sum_{j=1}^3 \beta_j \bU\check{\bsigma}_j\bU^\dagger.
    \label{bracket2}
    \ee
where
$\check\bsigma_j:=2\bsigma_j-\brho_d\bsigma_j\brho_d^{-1}-\brho_d^{-1}
\bsigma_j\brho_d$. In view of (\ref{eps=}), this relation implies
    \be
    \check\bsigma_j=-\frac{(\xi-\zeta)^2}{\xi\zeta}\times\left\{
    \begin{array}{ccc}
    \bsigma_1 & {\rm for} & j=1,\\
    \bsigma_2 & {\rm for} & j=2,\\
    \bzero & {\rm for} & j=3.
    \end{array}\right.
    \label{tilde-sigma}
    \ee

Next, we insert (\ref{tilde-sigma}) in (\ref{bracket2})
and use (\ref{UdU=}) to show that
    \bea
    [d\brho,\brho^{-1}]&=&-\frac{(\xi-\zeta)^2}{\xi\zeta}
    \bU\left(\sum_{j=1}^2 \beta_j\bsigma_j\right)\bU^\dagger
    =-\frac{(\xi-\zeta)^2}{\xi\zeta}
    \bU\left(\bU^\dagger d\bU-\beta_3\bsigma_3\right)\bU^\dagger\nn\\
    &=&-\frac{(\xi-\zeta)^2}{\xi\zeta}\left( d\bU\,\bU^\dagger-\beta_3
    \bU\,\bsigma_3\,\bU^\dagger\right).
    \label{bracket3}
    \eea
Because $\bU$ is a unitary matrix,
$d\bU\,\bU^\dagger=-\bU\,d\bU^\dagger$.  We also recall that
Hermitian-conjugation of $\bU$ has the same effect as changing
$\vartheta$ to $-\vartheta$. Therefore we can compute
$d\bU\,\bU^\dagger$ by taking $\vartheta$ to $-\vartheta$ in the
expression for $-\bU^\dagger\,d\bU$. In light of (\ref{UdU=}) and
(\ref{beta1}) -- (\ref{beta3}), this yields
$d\bU\,\bU^\dagger=\beta_1\bsigma_1+\beta_2\bsigma_2-\beta_3\bsigma_3$.
Substituting this equation and (\ref{U-id}) in (\ref{bracket3}), we
find
    \be
    [d\brho,\brho^{-1}]=-\frac{(\xi-\zeta)^2}{\xi\zeta}\left[(\beta_1-x_1\beta_3)\bsigma_1+
    (\beta_2-x_2\beta_3)\bsigma_2-(1+x_3)\beta_3\bsigma_3\right].
    \ee
With the help of this relation, we can use (\ref{hat-x=2}),
(\ref{bh-rho=}),  (\ref{tau-ttau-def}), and (\ref{beta1}) --
(\ref{beta3}) to establish (\ref{bh-rho=11}). Replacing
$(\xi,\zeta,\vartheta)$ with $(\tilde\xi,\tilde\zeta,\pi-\vartheta)$
on the right-hand side of (\ref{bh-rho=11}), we obtain
(\ref{tbh-rho=11}).

\ed